\documentclass[acmtog,pbalance]{acmart}

\definecolor{lightblue}{RGB}{240,245,255}
\definecolor{darkblue}{RGB}{40,40,85}

\usepackage[ruled]{algorithm2e} 

\SetAlFnt{\small}
\SetAlCapFnt{\small}
\SetAlCapNameFnt{\small}
\SetAlCapHSkip{0pt}
\usepackage{enumitem}
\setitemize{noitemsep,topsep=0pt,parsep=0pt,partopsep=0pt,leftmargin=*}

\newcommand{\dof}{\mathbf{q}}

\newcommand{\mat}{\mathbf}
\newcommand{\massmat}{\mathbf{M}}
\newcommand{\doftilde}{\widetilde{\mathbf{q}}}
\newcommand{\penaltyparam}{\rho}

\usepackage{graphicx}
\usepackage{xcolor}
\usepackage{makecell}
\usepackage{acmart-taps}
\newcommand{\changed}[2]{{#2}}
\acmSubmissionID{599}
\AtBeginDocument{%
  }

\copyrightyear{2026}
\acmYear{2026}
\setcopyright{cc}
\setcctype{by}
\acmConference[SIGGRAPH Conference Papers '26]{Special Interest Group on Computer Graphics and Interactive Techniques Conference Conference Papers}{July 19--23, 2026}{Los Angeles, CA, USA}
\acmBooktitle{Special Interest Group on Computer Graphics and Interactive Techniques Conference Conference Papers (SIGGRAPH Conference Papers '26), July 19--23, 2026, Los Angeles, CA, USA}
\acmDOI{10.1145/3799902.3811106}
\acmISBN{979-8-4007-2554-8/2026/07}

\begin{document}

%
%
%
%
%
%
%
%

\author{Jiafeng Liu}
\orcid{0000-0002-4962-5859}
\affiliation{\institution{State Key Lab of CAD and CG, Zhejiang University}
\city{Hangzhou}
\country{China}}
\email{jiafengliu@zju.edu.cn}

\author{Wenhui Zhou}
\orcid{0009-0009-8332-5102}
\affiliation{\institution{State Key Lab of CAD and CG, Zhejiang University}
\city{Hangzhou}
\country{China}}
\email{wenhuizhou@zju.edu.cn}

\author{Xinming Pei}
\orcid{0009-0001-3846-2986}
\affiliation{\institution{State Key Lab of CAD and CG, Zhejiang University}
\city{Hangzhou}
\country{China}}
\email{xmpei@zju.edu.cn}

\author{Yifan Peng}
\orcid{0000-0003-0667-2599}
\affiliation{\institution{The University of Hong Kong}
\city{Hong Kong}
\country{China}}
\email{evanpeng@hku.hk}

\author{Huamin Wang}
\orcid{0000-0002-8153-2337}
\affiliation{\institution{Style3D Research}
\city{Hangzhou}
\country{China}}
\email{wanghmin@gmail.com}

\author{Yin Yang}
\orcid{0000-0001-7645-5931}
\affiliation{\institution{University of Utah}
\city{Salt Lake City}
\country{USA}}
\email{yangzzzy@gmail.com}

\author{Lei Lan}
\orcid{0009-0002-7626-7580}
\affiliation{\institution{State Key Lab of CAD and CG, Zhejiang University}
\city{Hangzhou}
\country{China}}
\email{leilan@zju.edu.cn}
\authornote{Corresponding author.}

\author{Weiwei Xu}
\orcid{0000-0003-3756-3539}
\affiliation{\institution{State Key Lab of CAD and CG, Zhejiang University}
\city{Hangzhou}
\country{China}}
\email{xww@cad.zju.edu.cn}

\citestyle{acmauthoryear}
\renewcommand{\shortauthors}{J. Liu, W. Zhou, X. Pei, Y. Peng, H. Wang, Y. Yang, L. Lan, and W. Xu}

\title{Distributed Affine Body Dynamics with Adaptive Consensus}


\begin{abstract}

Affine Body Dynamics (ABD) within the Incremental Potential Contact (IPC) framework provides accurate simulation of extremely stiff solids exhibiting near-rigid behavior, with strict non-penetration guarantees. However, IPC’s globally coupled barrier constraints hinder scalable execution across multiple GPUs and compute nodes. We propose a distributed formulation of ABD using a consensus-based ADMM scheme. Each compute node solves its local ABD subproblem in parallel, followed by a global consensus step that enforces consistency among shared boundary bodies. The proposed method preserves IPC-level robustness and global consistency under distributed execution. Experiments demonstrate stable convergence, non-penetration, and efficient scaling on large-scale scenes across multiple nodes.

\end{abstract}




\begin{teaserfigure}
\includegraphics[width=\linewidth, trim=0px 0px 0px 0px,clip]{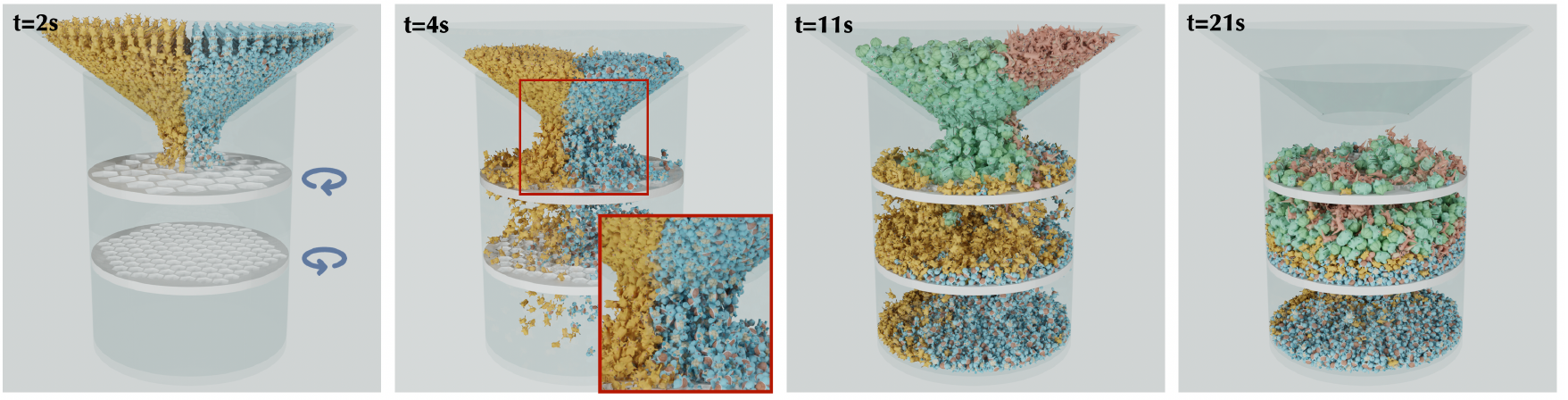} 
\vspace{-9pt}
\caption{A contact-rich simulation in which 5K ``Pok\'emon'' ABD bodies (6M triangles in total) fall into a narrow funnel under a two-worker distributed simulation setup. After entering the lower chamber, the bodies interact with two perforated turntables rotating in opposite directions, producing sustained stacking and recurring collisions. Our system remains penetration-free throughout the simulation despite complex geometry and cross-worker contacts.}
\label{fig:teaser}
\end{teaserfigure}


\maketitle


\section{Introduction}

Simulating materials that behave nearly rigid under contact is a recurring challenge in contact-rich physical interaction, particularly in embodied intelligence~\cite{long2025survey, barreiros2025learning}. In such settings, even minor violations of non-penetration can qualitatively alter interaction outcomes. The Incremental Potential Contact (IPC) framework~\cite{Li2020IPC} demonstrates that barrier-based formulations can robustly enforce non-penetration constraints under dense and dynamic contact. Within this framework, Affine Body Dynamics (ABD)~\cite{lan2022affine} provides a practical model for near-rigid materials by representing each body with 12 affine degrees of freedom, balancing numerical robustness and computational efficiency. Compared to conventional rigid body motion, ABD produces linear vertex trajectories over time, which simplifies continuous collision detection (CCD) in contact-rich scenarios.

Scaling ABD to large-scale scenes with dense contact remains
challenging. As scene complexity grows, the number of degrees of freedom increases rapidly, making single-device solvers inefficient due to the cost of assembling and solving large coupled systems. Dense contact configurations can generate tens to hundreds of millions of candidate collision pairs, incurring substantial overhead in collision detection and contact processing. While scaling ABD across parallel or distributed hardware is a natural way to overcome these limitations, in practice, barrier-based contact formulations introduce strong global coupling across interacting objects, causing naive domain decomposition and local solve strategies to fail.

%
In light of these challenges, we propose a distributed simulation framework for ABD that runs across multiple compute nodes in parallel and maintains penetration-free contact throughout the simulation. 
In contact-rich scenarios, IPC contact constraints introduce strong coupling among interacting bodies. As a result, distributed ABD simulation constitutes a \emph{consensus optimization} problem in which consistency must be enforced on the states of bodies shared across partitions to ensure a well-defined global solution.

We solve this optimization problem using the Alternating Direction Method of Multipliers (ADMM), which decomposes a global optimization problem into local subproblems and enforces consistency among them through augmented Lagrangian updates. 
The convergence of ADMM is known to be sensitive to the choice of the penalty parameter~$\rho$, particularly in strongly coupled distributed problems.
To address this challenge, we introduce an adaptive consensus mechanism that regulates the strength of consistency enforcement by adaptively adjusting the penalty parameter~$\rho$. We adopt a mass-aware strategy to initialize~$\rho$ based on body masses, and further refine it through residual-driven adaptation during the iterations. Mass-aware initialization establishes a stable baseline, while adaptive consensus dynamically maintains balanced progress among heterogeneous subproblems, resulting in faster and more stable convergence across partitions.

In addition, ensuring non-penetration in contact-rich distributed simulation requires particular care. 
Even if local updates remain penetration-free under the IPC barrier energy, naive synchronization across partitions can still introduce interpenetration. 
To address this, we introduce a \emph{feasibility-preserving consensus} mechanism that combines locally valid updates with a global, CCD-certified consensus step, applying synchronized states only when they pass the CCD check.
This supports stable, penetration-free simulation at large scale, handling thousands of rigid bodies with surface meshes containing tens of millions of triangles, while maintaining strong parallel scalability in distributed environments.

In summary, this work makes the following contributions:
\begin{itemize}[leftmargin=*,nosep]
  \item We present a distributed ABD framework based on ADMM that preserves IPC-style non-penetration guarantees while enabling scalable execution across multiple GPUs and compute nodes for large-scale, contact-rich simulations.
  \item We propose a mass-aware initialization of the ADMM penalty parameter, combined with residual-driven adaptation, which accelerates convergence and improves robustness in distributed ABD optimization.
  \item We demonstrate the effectiveness of the proposed method on large-scale, contact-rich simulation scenarios, showing stable convergence, penetration-free simulation, and efficient scaling in distributed environments.
\end{itemize}



\section{Related Work}
\paragraph{ADMM for simulation} ADMM is widely used in physics-based simulation. By decomposing a global optimization into simple, parallelizable subproblems, it enables efficient solvers for fluids~\cite{inglis2017primal, pan2017efficient}, elastic bodies~\cite{overby2017admm, fang2019silly, brown2021wrapd, daviet2023interactive}, multibody systems~\cite{lee2023modular, ji2025gpu}, and contact handling~\cite{daviet2020simple, tasora2021solving, FrictionShockChen23}.
However, ADMM often slows down near convergence. Recent work accelerates ADMM by viewing it as a fixed-point iteration and applying Anderson acceleration or Douglas-Rachford splitting (DRS)-based reformulations~\cite{peng2018Anderson,Zhang2019Accelerating,ouyang2020anderson}.
ADMM is also sensitive to the penalty parameter. For deformable simulation, prior methods tune it using stiffness/curvature proxies to balance local and global updates~\cite{overby2017admm, brown2021wrapd, daviet2020simple, fang2019silly}. This strategy is less effective in our distributed ABD setting. We instead adopt a mass-aware penalty with residual-driven updates, which substantially improves convergence in our system.

\paragraph{Penetration-free contact handling}
Penetration-free contact is essential for physics-based simulation, as interpenetration undermines plausibility and can destabilize contact response.
Incremental Potential Contact (IPC)~\cite{Li2020IPC} is a robust interior-point framework that maintains a feasible optimization path using CCD-based line search and a barrier contact energy.
IPC has been extended to cloth~\cite{li2020codimensional,li2023subspace, lan2024efficient}, rigid (or close-to-rigid) bodies~\cite{ferguson2021intersection, lan2022affine, chen2022unified}, and fluid-rigid coupling~\cite{xie2023contact}, but its computational cost has motivated many acceleration efforts.
Recent GPU-oriented advances include coordinate descent~\cite{lan2023second}, Gauss--Newton Hessian approximations~\cite{huang2024gipc}, MAS-preconditioned ABD-IPC~\cite{huang2024stiffgipc}, and barrier-augmented Lagrangian variants~\cite{guo2024barrier}. Besides, non-barrier alternatives also provide efficient solutions~\cite{wang2023fast, lan2024efficient, zheng2025robust, chen2025offset}.
Local-global methods~\cite{lan2022penetration, li2023subspace} are particularly relevant due to their alternating structure.
PD-IPC~\cite{lan2022penetration} replaces nonlinear barrier iterations with massively parallel local projections and a global linear solve.
Our method adopts a similar local/merge alternation, but treats the ADMM consensus state as an intermediate estimate and enforces penetration-freeness only at termination.

\paragraph{Distributed optimization} \changed{}{Prior research on distributed optimization spans both optimization methods, such as decomposition ~\cite{palomar2006tutorial, chan1994domain} and consensus-based optimization schemes~\cite{nedic2009distributed,shi2015extra,yuan2018exact,boyd2011distributed,yang2022survey}, and system frameworks for large-scale distributed execution~\cite{dean2008mapreduce,gropp2001mpi,fetterly2009dryadlinq,zaharia2012resilient,low2012distributed}. These ideas naturally connect to distributed simulation, where physical systems are partitioned across devices and coordinated through cross-partition communication.} \changed{Distributed simulation scales physics-based computation by partitioning work across multiple GPUs or machines. Prior work covers cloth, fluids, MPM, and rigid bodies.}{} P-cloth~\cite{pcloth20} distributes the dominant linear solve for implicit cloth stepping across multiple GPUs. Rustico et al.~\shortcite{rustico2012advances} propose a distributed SPH fluid solver with dynamic load balancing, and Liu et al.~\shortcite{liu2016scalable} develop a scalable Eulerian method via a Schur-complement formulation. \emph{Nimbus} further automates distributed execution for grid-based fluids~\cite{, mashayekhi2018automatically}. Wang et al.~\shortcite{wang2020massively} exchange the halo-region data of MPM between GPU partitions. Qiu et al.~\shortcite{qiu2023sparse} extend this line with sparse grids on GPU-equipped supercomputers. Brown et al.~\shortcite{brown2020distributed} use aura-based migration to co-locate potentially colliding rigid bodies, and \emph{Loki}~\cite{lesser2022loki} targets multiphysics parallel execution \changed{on a single multi-core CPU node}{across multiple CPU compute nodes.}

\changed{}{ Prior work has explored split-and-merge style strategies for parallel and distributed simulation. Kale and Kry~\shortcite{kale2023distributed} use overlapping interface-body sets with state blending. Tonge et al.~\shortcite{tonge2012mass} introduce mass splitting, and Daviet ~\shortcite{daviet2023interactive} applies a similar split-and-average idea to GPU-parallel contact projection across multiple GPUs. These works are closely related to our method in that they all rely on duplicating interface states and reconciling them through a merge step. However, they do not address high-accuracy collision response with strict non-penetration enforcement in contact-rich distributed simulation.}



\section{Preliminaries}


\paragraph{Implicit Euler integration via optimization.}
The backbone formulation of our framework follows ABD, which advances the simulation by solving the following variational problem:
\begin{equation}
    \dof^{t+1} = \arg \min_{\dof} \frac{1}{2} \big\|\dof - \widetilde{\dof}\big\|_{\massmat}^2 + h^2E_{ARAP}(\dof) + h^2E_{IPC}(\dof),
\end{equation}
where $\dof$ denotes the generalized coordinates of the system, $\massmat$ is the mass matrix of affine bodies, $E_{ARAP}$ and $E_{IPC}$ are the elastic energy and collision energies, respectively. 
$\widetilde{\dof} = \dof^t + h \dot{\dof} + h^2 {\massmat}^{-1} \mat{f}^{t+1}_{ext}$ is the predicted position 
based on the inertia and external forces, where $h$ is the time step and $\mat f_{ext}$ is the external generalized force. For detailed definitions of the DoFs and energy terms, we refer to the prior work~\cite{lan2022affine}.

\begin{figure}[t]
  \centering
  \includegraphics[width=\columnwidth]{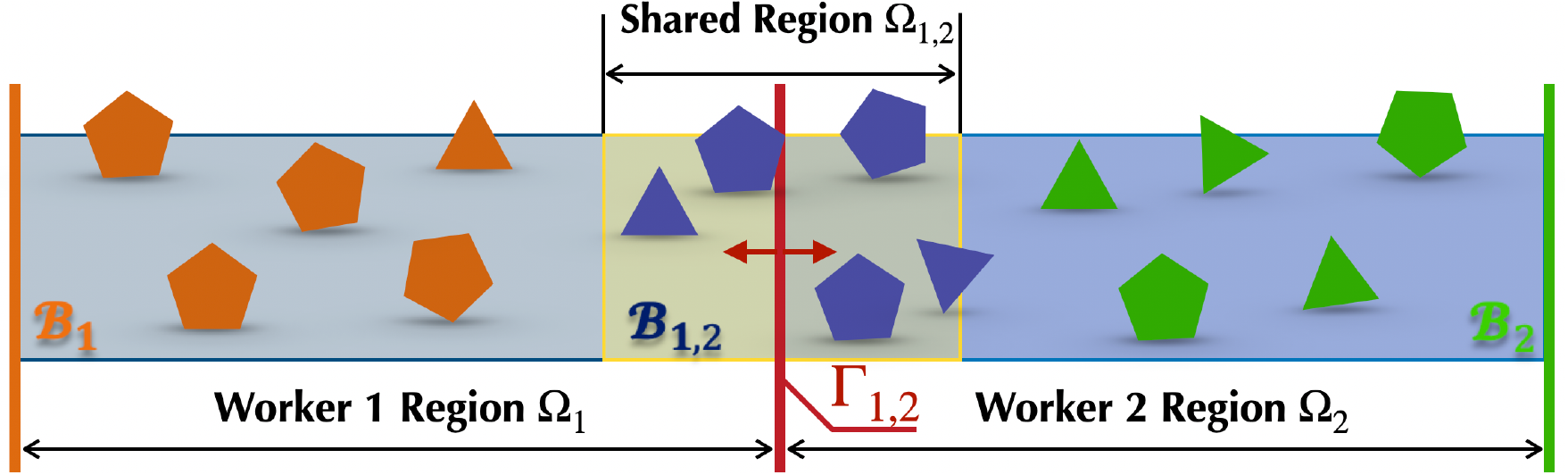}
  \vspace{-9pt}
  \caption{Partition model. 
  The entire scene is partitioned into two regions ($\Omega_1, \Omega_2$) by the boundary $\Gamma_{1,2}$. We set up an overlap area as \emph{Shared Region } $\Omega_{1,2}$ around $\Gamma_{1,2}$.
  Objects with the same color belong to the same set. Orange and green denote the sets of internal bodies of worker 1 ($\mathcal{B}_1$) and worker 2 ($\mathcal{B}_2$), respectively, while blue indicates the geometry of the \emph{shared bodies}. Note that the partition boundary $\Gamma_{1,2}$ is a plane that can only translate along its normal direction, as indicated by the red arrows in the figure.
  }
  \vspace{-9pt}
  \label{fig:partition_model}
\end{figure}


\section{Distributed Simulation with Consensus}
For clarity, we first introduce the distributed formulation.
Fig. ~\ref{fig:partition_model} provides a concrete example of two workers.
We consider a distributed system with $N$ workers, indexed by $\mathcal{W}:=\{1,\dots,N\}$. Let $\mathcal{B} := \{b_1,b_2,\dots,b_B\}$ denote the set of bodies in the scene, where $B=|\mathcal{B}|$. 
Each worker communicates only with its neighboring workers. For worker $i$, let $\mathcal{N}_i$ denote its neighbor set.
By definition, $i\notin\mathcal{N}_i$, and the neighbor relation is symmetric: $j\in\mathcal{N}_i \Leftrightarrow   i\in\mathcal{N}_j$. We partition the scene into sub-regions ${\{\Omega_i\}}$ and assign worker $i$ to region $\Omega_i$. For neighboring workers $i$ and $j$ (with $j\in\mathcal{N}_i$), we denote their interface by $\Gamma_{ij}$ and introduce a shared overlap region $\Omega_{ij}$ to capture cross-interface interaction. Under this model, worker $i$ maintains two collections of bodies: $\mathcal{B}_i$, the internal bodies of worker $i$, and $\mathcal{B}_{ij}$, the bodies shared with neighbor $j$. These shared bodies constitute the interface variables whose states are duplicated across workers and coordinated through consensus in the distributed formulation. \changed{}{For the overlap-region width, we use $w=\max(\alpha v_{\max}dt, w_{\min})$ with $\alpha=2$, where $v_{\max}$ is the maximum body velocity and $w_{\min}$ is the smallest body-bounding-box diagonal. The first term prevents missed bodies across one time step, and the second prevents an overly narrow shared region at low velocity.}

\subsection{Consensus ADMM Solver}

\paragraph{Problem definition.}
We now define the optimization problem solved at each time step using ADMM.
The global ABD system is partitioned across $N$ workers and solved in a distributed manner, which naturally gives rise to a \emph{consensus optimization} problem:
\begin{equation}
\label{equ:global_obj}
 \min_{\{\dof_i\},\{\mat z_b\}}\; \sum_{i=1}^{N} F_i(\dof_i), \;
 \text{s.t.}\; \dof_{i,b} - \mat z_b = 0,\; \forall i=1,\dots,N,\; \forall b\in\mathcal{B}_{i,\partial}.
\end{equation}

where $\mathcal{B}_{i,\partial} := \bigcup_{j \in \mathcal{N}_i} \mathcal{B}_{ij}$ denotes the set of shared bodies handled by worker $i$, and $\{\mathbf z_b\}_{b \in \mathcal{B}_{\partial}}$ are the consensus variables, with $\mathcal{B}_{\partial} := \bigcup_{i=1}^N \mathcal{B}_{i,\partial}$.
The vector $\dof_i$ stacks the local DoFs maintained by worker $i$, including replicas of shared bodies, and $\dof_{i,b}$ denotes the DoFs block corresponding to body $b$ within $\dof_i$.
A naive summation $\sum_i F_i$ would over-count contributions associated with replicated bodies or cross-partition interactions, which motivates the consensus formulation.

To ensure consistency of the global objective with the single-worker formulation, we normalize these contributions by their replication counts. Specifically, let $\kappa_b$ denote the number of workers that hold a replica of body $b$, and let $\kappa_c$ denote the number of workers that include a contact pair $c$.

The local objective on worker $i$ is defined as
\begin{align}
F_i(\dof_i)
=&\sum_{b\in \mathcal{B}_i\cup\mathcal{B}_{i,\partial}} \frac{1}{\kappa_b}
\Big(\frac{1}{2}\big\|\dof_{i,b}-\doftilde_{i,b}\big\|_{\massmat_{i,b}}^2
+h^2E_{\mathrm{ARAP}}(\dof_{i,b})\Big) \notag \\
&+h^2\sum_{c\in\mathcal{C}_i}\frac{1}{\kappa_c}\,\phi_c(\dof_i) \, ,
\end{align}
where $\mathcal{C}_i$ is the set of IPC contact pairs on worker $i$, and $\phi_c(\cdot)$ is the IPC barrier term for pair $c$. This construction ensures that each per-body and contact-pair term contributes exactly once to the global objective $\sum_i F_i$.

We apply ADMM to Eq.~\eqref{equ:global_obj} by constructing the augmented Lagrangian in scaled form~\cite{boyd2011distributed}:
\begin{equation}
\label{equ:alm}
\changed{}{
\mathcal{L}}
=\sum_{i=1}^{N}\left(
F_i(\dof_i)+\sum_{b\in\mathcal{B}_{i,\partial}}
\left[
\frac{\rho_{i,b}}{2}\big\|\dof_{i,b}-\mat z_b+\mat u_{i,b}\big\|^2\changed{}{
-\frac{\rho_{i,b}}{2}\|\mat u_{i,b}\|^2}
\right]
\right).
\end{equation}
where \changed{}{$\mathcal{L}$ denotes the scaled-form augmented Lagrangian, and we suppress its arguments for brevity}, $\mathbf u_{i,b}$ denotes the scaled dual variable and $\rho_{i,b}$ is the penalty parameter.
Starting from initial values $\{\dof_i^{0}\}$, $\{\mathbf z_b^{0}\}$, and $\{\mathbf u_{i,b}^{0}\}$, ADMM alternates between local solves, consensus updates, and dual updates. 

At iteration $k$, each worker $i$ performs a local solve by minimizing $F_i$ with respect to $\dof_i$, while softly anchoring each shared body block $\dof_{i,b}$ to the current consensus $\mathbf z_b^{k}$ via a quadratic penalty weighted by $\rho_{i,b}$:
\begin{equation}\label{equ:local_solve}
\dof_i^{k+1}
=\arg\min_{\dof_i}\
F_i(\dof_i)+\sum_{b\in\mathcal{B}_{i,\partial}}
\frac{\rho_{i,b}}{2}\left\|\dof_{i,b}-\mat z_b^{k}+\mat u_{i,b}^{k}\right\|^2.
\end{equation}

A consensus step is then performed for each shared body $b$ by aggregating the updated local blocks
$\{\dof_{i,b}^{k+1}+\mat u_{i,b}^{k}\}$ for $i\in\mathcal{W}_b$ across workers to obtain the new consensus variable $\mathbf z_b^{k+1}$:
\begin{equation}\label{equ:consensus_solve}
\mat z_b^{k+1}
=\arg\min_{\mat z_b}\ \sum_{i\in\mathcal{W}_b}
\frac{\rho_{i,b}}{2}\left\|\mat z_b-\left(\dof_{i,b}^{k+1}+\mat u_{i,b}^{k}\right)\right\|^2.
\end{equation}

Here, $\mathcal{W}_b := \{\, i \in \mathcal{W} \mid b \in \mathcal{B}_{i,\partial} \,\}$ denotes the set of workers that store a local copy of body $b$.
The dual update then adjusts the scaled multipliers $\mathbf u_{i,b}$ based on the current primal residual, gradually reducing discrepancies among replicated variables:
\begin{equation}\label{equ:update_dual}
\mat u_{i,b}^{k+1}
=\mat u_{i,b}^{k}+\dof_{i,b}^{k+1}-\mat z_b^{k+1}.
\end{equation}
Note that the $\mathbf z$ update does not require a centralized worker.
For each shared body $b$, every worker in $\mathcal{W}_b$ can compute the same $\mathbf z_b^{k+1}$ locally by exchanging
$\{\dof_{j,b}^{k+1}, \mathbf u_{j,b}^{k}, \rho_{j,b}\}$ with neighboring workers that hold replicas of $b$.

\paragraph{Stopping criteria.}
ADMM iterations are terminated based on three metrics:
the Newton update magnitude $\|\Delta  \dof_i^{k+1}\|_\infty$,
the primal residual $r_\infty^{k+1}$,
and the dual residual $s_\infty^{k+1}$,
all normalized by the scene scale $l$ and evaluated using the $\ell_\infty$ norm:
\begin{equation}\label{equ:criteria}
\max_{i=1,\dots,N}\frac{1}{h\,l}\,\|\Delta \dof_i^{k+1}\|_{\infty} < \theta,
\quad
\frac{1}{h\,l}r_{\infty}^{k+1} < \theta,
\quad
\frac{1}{h\,l}s_{\infty}^{k+1} < \theta.
\end{equation}
Here, $\theta$ is a threshold on displacement changes applied to all three metrics.
The primal residual captures the worst-case consensus mismatch on shared bodies and is defined as
\begin{equation}\label{equ:compute_pr}
\mathbf r_{i,b}^{k+1} := \dof_{i,b}^{k+1} - \mathbf z_b^{k+1},\quad
r_\infty^{k+1} := \max_{b\in\mathcal{B}_\partial}\ \max_{i\in\mathcal{W}_b}\ \|\mathbf r_{i,b}^{k+1}\|_{\infty}.
\end{equation}
The dual residual measures the change of the merged consensus state:
\begin{equation}\label{equ:compute_dr}
\mathbf s_b^{k+1} := \mathbf z_b^{k+1}-\mathbf z_b^{k},\qquad
s_\infty^{k+1} := \max_{b\in\mathcal{B}_\partial}\ \|\mathbf s_b^{k+1}\|_{\infty}.
\end{equation}
A penetration-free check is enforced to ensure feasibility of the synchronized state; see Sec.~\ref{sec:penetration-free} for details. After convergence, shared bodies are merged by replacing all local replicas $\dof_{i,b}$ with the corresponding consensus states $\mathbf z_b$.

\subsection{Penetration-Free Consensus Update}
\label{sec:penetration-free}


\begin{figure}[t]
  \centering
  \includegraphics[width=0.78\columnwidth]{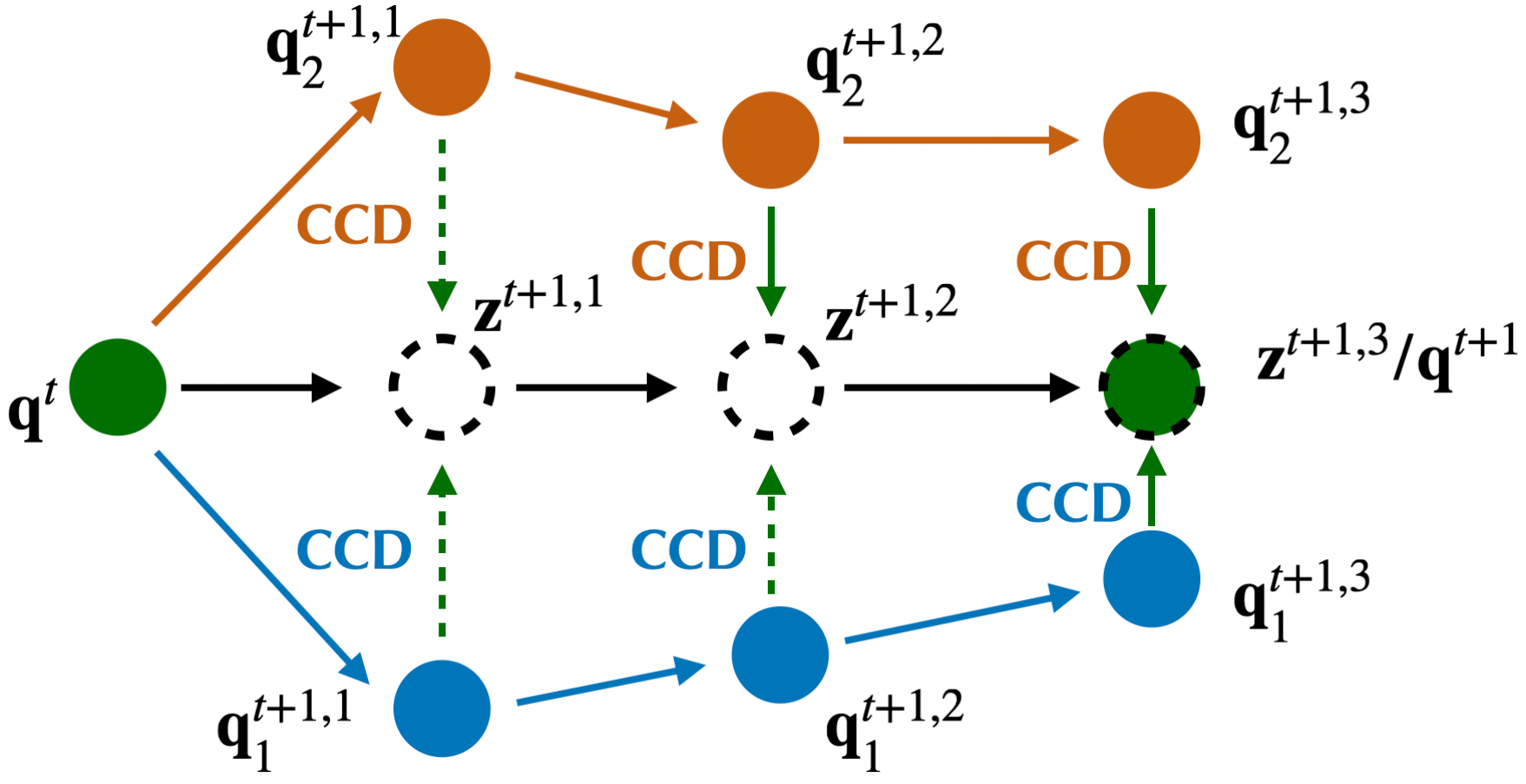}
  \vspace{-6pt}
  \caption{Illustration of our CCD check process.  We show one shared body evolving from $\mathbf{q}^{t}$ to $\mathbf{q}^{t+1}$ in a two-worker solve, where blue/orange denote the local copies on worker 1/2. Each ADMM iteration updates $\mathbf{q}_i^{t+1,k}$ by a local solve and then forms the consensus $\mathbf{z}^k$. To ensure a penetration-free step, we run CCD from each $\mathbf{q}_i^{t+1,k}$ toward $\mathbf{z}^k$. In the figure, green dashed lines indicate failed checks ($\mathrm{TOI}\neq 1$), while green solid lines indicate passed checks ($\mathrm{TOI}=1$). We accept $\mathbf{q}^{t+1}=\mathbf{z}^k$ only when all workers report $\mathrm{TOI}=1$.
}\vspace{-6pt}
  \label{fig:global_ccd}
\end{figure}


In our formulation, each worker evolves its local configuration $\dof_i$, while $\mathbf z$ serves only as an auxiliary consensus variable in ADMM.
Each local solve produces a penetration-free $\dof_i$, but $\mathbf z$ may be interpenetrating. As a result, even after ADMM has converged, applying the consensus state to shared bodies in the final merge can still introduce interpenetration.

Based on this consideration, we augment the ADMM convergence test with an additional CCD check on the merge target, as illustrated in Fig.~\ref{fig:global_ccd}.
At iteration $k$, the local solve yields a penetration-free $\dof_i^{k+1}$, while the consensus step produces the shared-body estimate $\mathbf z^{k+1}$.
A candidate merged state $\hat{\dof}_i^{k+1}$ is constructed by replacing every shared body $b \in \mathcal{B}_{i,\partial}$ in $\dof_i^{k+1}$ with $\mathbf z_b^{k+1}$.
CCD is then performed along the trajectory from $\dof_i^{k+1}$ to $\hat{\dof}_i^{k+1}$.
Convergence is accepted only if this check passes on every worker, ensuring that the consensus state can be safely applied without introducing penetration.

\paragraph{Discussion}
\changed{ A similar issue also arises in PD-IPC~\cite{lan2022penetration},  which follows an alternating local-global (L-G) iteration scheme. Its local and global steps are  analogous to our local update in Eq.~\eqref{equ:local_solve} and consensus update in Eq.~\eqref{equ:consensus_solve}. The local step solves a set of subproblems to obtain local projection targets. The global step then combines these projections by solving a global linear system. The resulting consensus variable $\mathbf z$ is taken as the global configuration for the next iteration. Therefore, after each L-G iteration, CCD is applied toward $\mathbf z$ to ensure that it remains penetration-free, by truncating the update at the time of impact ($\mathrm{TOI}$).
}{}

\changed{
In our distributed setting, the consensus variable $\mathbf z$ is used differently from that in PD-IPC. It serves as an auxiliary estimate during ADMM iterations and is not required to remain penetration-free at intermediate steps. Enforcing penetration-free on every intermediate $\mathbf z$, as in PD-IPC, would be overly conservative in a distributed system.
Early in the ADMM iteration, replicas of shared bodies can be far from agreement, causing CCD to return a very small $\mathrm{TOI}$ and severely restrict the consensus update, which in turn slows residual reduction.
Moreover, using the global minimum $\mathrm{TOI}$ forces all workers to follow the most restrictive constraint, further limiting parallel progress.
Instead, we apply the CCD check only as part of the convergence test to ensure that the final merged configuration is penetration-free.
Empirically, this strategy is sufficient in our experiments, and we observe no interpenetration after merging.
}{}

\begin{figure}[t]
\centering
  \includegraphics[width=0.95\columnwidth]{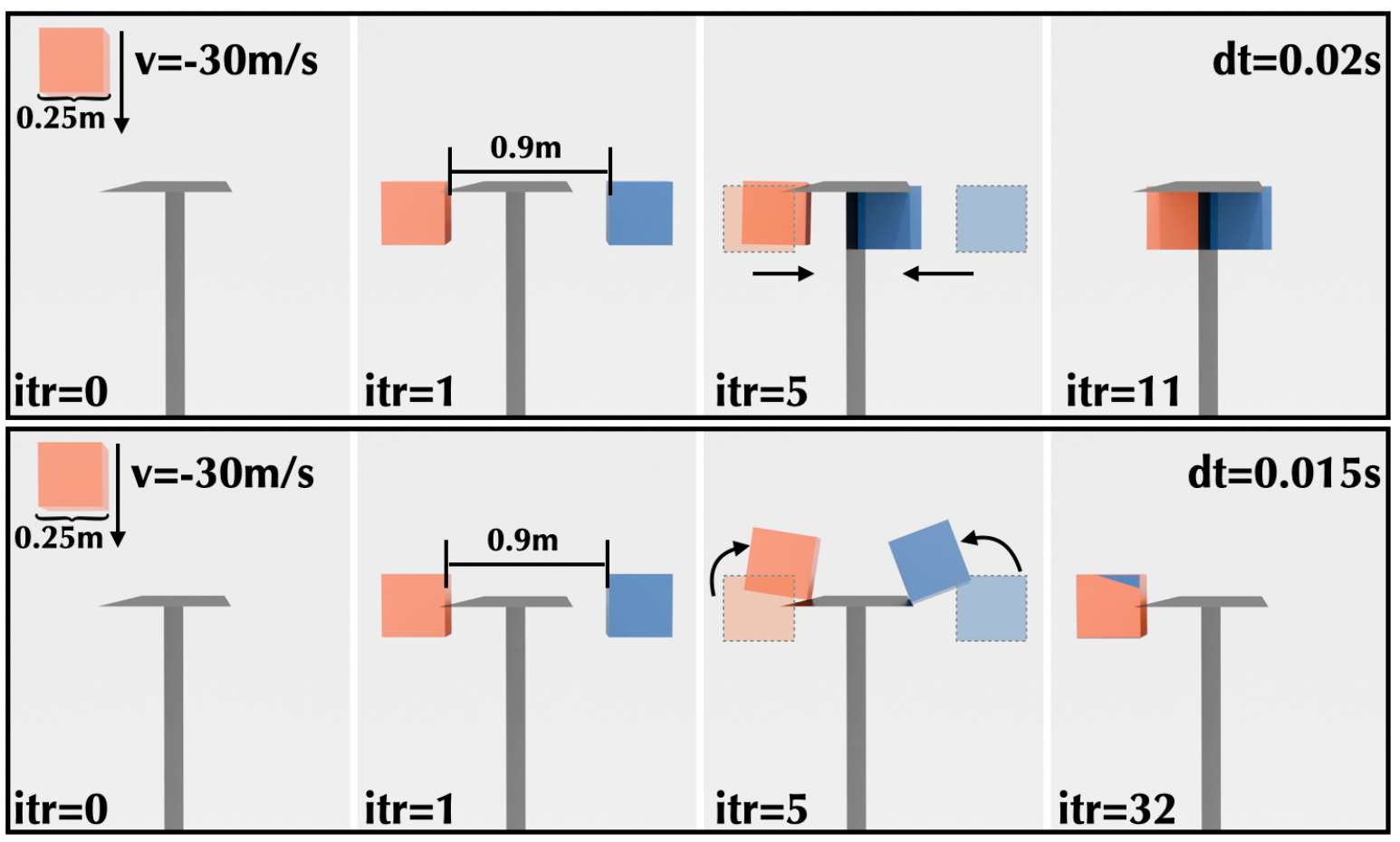}
    \caption{\changed{}{A CCD-check corner case with a fixed blocking obstacle. In both rows, the duplicated body starts with the same downward velocity ($v=-30\,\mathrm{m/s}$) and the same replica separation ($0.9\,\mathrm{m}$) at $itr=1$. The dashed silhouettes indicate the replica positions at $itr=1$, and the arrows illustrate their subsequent motion directions during ADMM iterations. With $dt=0.02\,\mathrm{s}$ (top), the merge path is blocked and the replicas cannot merge. Reducing the time step to $dt=0.015\,\mathrm{s}$ (bottom) alleviates this issue and allows the replicas to gradually move around the obstacle and eventually merge.}}
\label{fig:exp-ccd-cornercase-exp}
\end{figure}

\changed{}{With CCD enforced, a natural concern is whether repeated CCD failure could hinder convergence. We observe this issue is effectively mitigated by an adaptive time-stepping strategy. To examine this corner case, we construct the two-worker example in Fig.~\ref{fig:exp-ccd-cornercase-exp}, where two replicas of the same body are driven onto opposite sides of an obstacle using a large velocity and time step. This configuration is highly artificial, requiring a horizontal force on the order of 586$\times$ the body’s weight. In this case, the obstacle blocks the direct merge path. Reducing the step size allows the replicas to move around the obstacle and eventually merge. Across all other experiments, we observe no CCD-induced non-convergence.}

\subsection{Mass-aware Penalty Parameter Adaptation} 
The penalty parameter $\rho$ plays a critical role in the convergence of ADMM. 
When $\rho$ is too small, the consensus penalty is too weak to effectively reduce discrepancies among split degrees of freedom, resulting in slow agreement across partitions. Conversely, when $\rho$ is too large, the quadratic penalty dominates the local subproblems, pulling updates aggressively toward the consensus variables and limiting progress on the original objective.
We initialize the ADMM penalty parameter $\rho$ using a mass-aware strategy, and further refine it through residual-driven adaptation during the iterations. 

\paragraph{Mass-aware stiffness estimation} We set the initial value of $\rho$ by $\rho_{i,b} = \beta m_b$, where $\beta$ is a user-specified factor. Unless stated otherwise, we set $\beta=1$ in all experiments.
A common strategy in prior ADMM-based solvers is to scale the penalty parameter using stiffness-related curvature proxies, such as eigenvalue-based statistics of local Hessian/constraint blocks~\cite{overby2017admm, brown2021wrapd, fang2019silly}.
Following this strategy, for a body $b$ on worker $i$ (with DoFs $\dof_{i,b}$), one may analyze the stiffness of the Hessian diagonal block
\begin{align}
\mathbf{H}_{i,b}
=\mathbf{M}_{i,b}
+\mathbf{H}^{\mathrm{ARAP}}_{i,b}
+\mathbf{H}^{\mathrm{IPC}}_{i,b},
\end{align}
where $\mathbf{M}_{i,b}$, $\mathbf{H}^{\mathrm{ARAP}}_{i,b}$, and $\mathbf{H}^{\mathrm{IPC}}_{i,b}$ denote the contributions from inertia, elasticity, and the IPC barrier, respectively.
However, directly conducting eigen analysis, e.g. using $\lambda_{\max}(\mathbf{H}_{i,b})$ or the average of $\lambda(\mathbf{H}_{i,b})$ can severely overestimate the effective stiffness, because the largest eigenvalues of $\mathbf{H}_{i,b}$ are often induced by only a few constrained modes in the ARAP and IPC terms.

Specifically, for ARAP, the largest eigenvalues mainly correspond to the three scaling modes, and can reach $10^{8}$ in our experiments (given a Young's modulus of $10^{12}$), which easily dominates $\lambda_{\max}(\mathbf{H}_{i,b})$.
Besides, the IPC term is a log-barrier function. When two bodies are very close, its associated eigenvalues can become extremely large.
Moreover, this barrier only penalizes the relative-motion directions that change $d$, so the resulting large curvature can again dominate $\lambda_{\max}(\mathbf{H}_{i,b})$ and yield an overly conservative stiffness proxy~\cite{daviet2020simple}. In addition, barrier pairs may occasionally activate and deactivate across iterations as distances fluctuate around the activation threshold $\hat d$, making the curvature contribution of $\mathbf{H}^{\mathrm{IPC}}_{i,b}$ 
 highly transient.
Therefore, we exclude the ARAP and IPC terms in stiffness estimation and rely only on the inertial block $\mathbf{M}_{i,b}$. 
In the ABD formulation, $\mathbf{M}_{i,b}$ has a simple spectral bound: its largest eigenvalue satisfies $\lambda_{\max}(\mathbf{M}_{i,b}) \le m_b$, so we use the body mass $m_b$ as a simple stiffness proxy. The derivation is provided in the supplementary material.

\paragraph{Residual-driven adaptation of $\penaltyparam$.}

\begin{figure}[t]
  \centering
  \includegraphics[width=0.85\columnwidth, trim=5pt 0pt 0pt 40pt,clip]{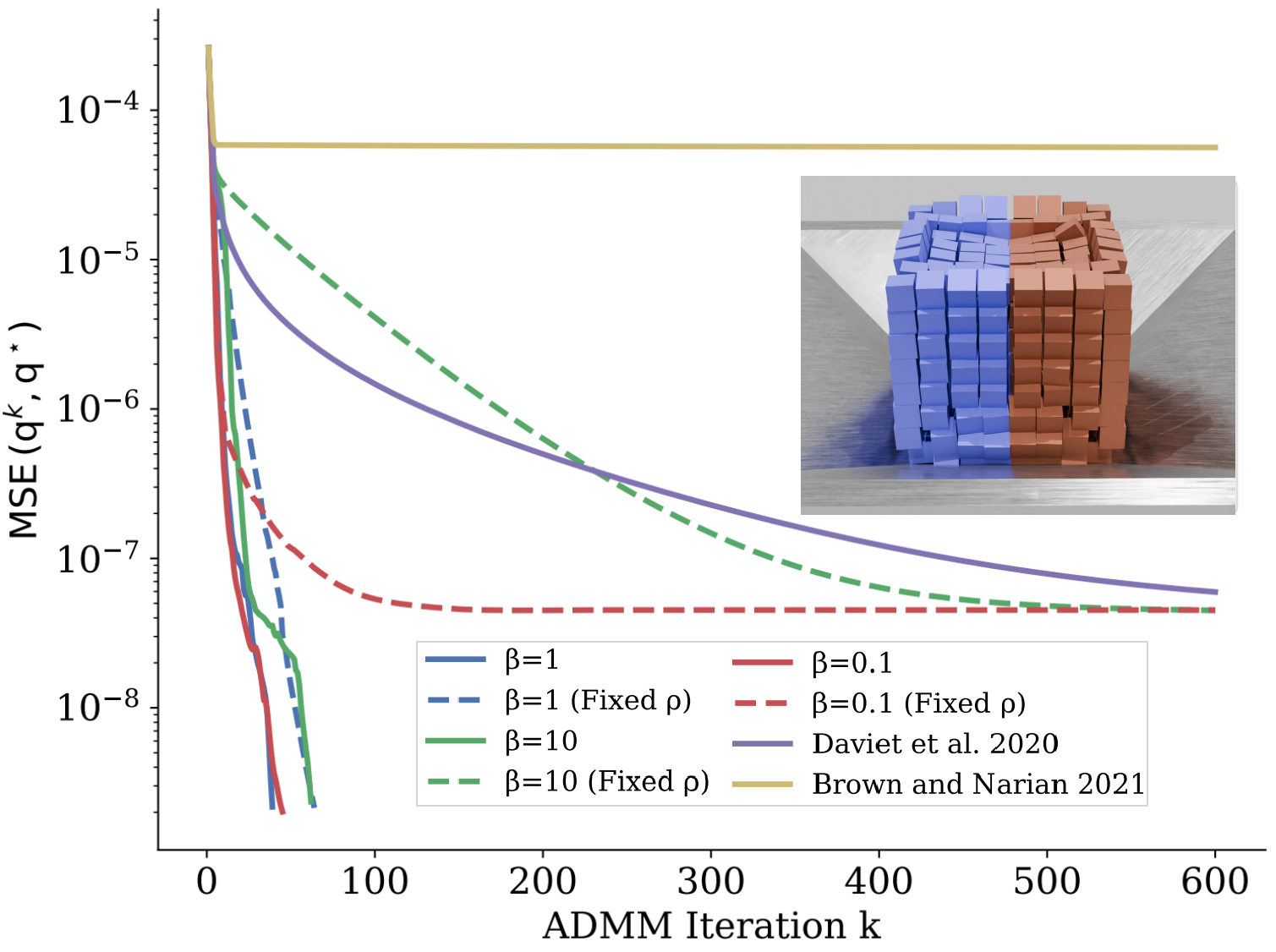}
  \caption{Convergence comparison of $\rho$ adaptation methods. MSE to a single-worker reference solution $\mathbf{q}^\star$ versus ADMM iterations, comparing our update to prior rules~\cite{daviet2020simple,brown2021wrapd}.}
  \label{fig:residual}
\end{figure}

In contact-rich distributed ABD simulation, although the consensus requirement itself remains unchanged, the appropriate enforcement strength varies as contacts activate, deactivate, and redistribute coupling across partitions.
To account for this effect, we introduce an \emph{adaptive consensus} mechanism that dynamically regulates the ADMM penalty parameter $\rho$ via residual-driven, per-body adaptation, allowing the consensus strength to respond to heterogeneous convergence rate induced by varying contact states and coupling intensity. Following the residual-driven adaptation scheme of Boyd et al. ~\shortcite{boyd2011distributed}, we update the per-body penalty parameter $\rho_b$ as

\begin{equation}
\penaltyparam_b^{k+1} =
\begin{cases}
\tau\,\penaltyparam_b^{k}, & \|\mat r_b^{k}\|_{\infty} > \mu \|\mat s_b^{k}\|_{\infty}, \\
\penaltyparam_b^{k}/\tau, & \|\mat s_b^{k}\|_{\infty} > \mu \|\mat r_b^{k}\|_{\infty}, \\
\penaltyparam_b^{k}, & \text{otherwise}.
\end{cases}
\end{equation}
where $\mathbf r_b^k$ and $\mathbf s_b^k$ denote the primal and dual residuals of body $b$ at iteration $k$.
By comparing these residuals, $\rho_b$ is adjusted to balance global consistency and local objective progress.
We employ multiplicative updates with threshold $\mu$ for rapid imbalance correction while avoiding sensitivity to residual noise.
For numerical stability, $\rho_b$ is clamped to a bounded range around its initial mass-aware value.
Unless otherwise stated, we use \changed{$\tau=5$, $\mu=2$}{$\tau=2$, $\mu=5$}, and clamp $\rho_b$ to $[\sigma_{\min}\penaltyparam^0,\sigma_{\max}\penaltyparam^0]$. In our experiments, we set $\sigma_{\min}=0.001$ and $\sigma_{\max}=1000$ .

We evaluate our penalty-parameter strategy in a two-worker scene with 512 ABD bodies shown in Fig.~\ref{fig:residual}, comparing against the dynamic-penalty methods of~\cite{daviet2020simple, brown2021wrapd}. A non-distributed ABD solution serves as the reference state $\dof^\star$. We then run the same scene on two nodes and report, at the $k$-th ADMM iteration, the mean squared error (MSE) between $\dof^{k}$ and $\dof^\star$.
 Our method rapidly reduces the error and reaches near-reference accuracy within 80 iterations. In contrast,~\cite{brown2021wrapd} shows little progress beyond the initial drop due to setting $\rho$ based on a very large ARAP stiffness, while~\cite{daviet2020simple} converges faster but remains noticeably slower than our approach. We further validate the effectiveness of residual-driven adaptation: for both $\beta=10$ and $\beta=0.1$, enabling this update consistently converges faster than a fixed $\rho$ and drives the solution closer to $\dof^\star$.




\section{Implementation Details}
\subsection{Distributed System Overview}
\paragraph{Overview.} Our system adopts a single-controller, multi-worker architecture, where the controller orchestrates the simulation loop and global convergence via frame-level commands. Workers are GPU-equipped nodes that perform local Newton solves and communicate only with the controller and neighboring workers under the partition model in Fig.~\ref{fig:partition_model}. \changed{The full execution pipeline is provided in the supplementary material.}{
We summarize the system pipeline in Algorithm~\ref{algorithm:worker_consensus_admm}. At the beginning of each frame, worker~$i$ fetches the shared-body data from its neighbors, builds the local frame data, and initializes $\mat z_i^1=\widetilde{\dof}_i$ and $\mat u_i^1=\mat 0$ (Lines~1--3). The fetched data include the mass matrices, mesh identifiers, and current states ${\dof_b,\dot{\dof}_b}$ of the shared bodies. The worker then enters the ADMM loop (Lines~4--14). For $k>1$, it fetches the current shared-body quantities ${\dof_{j,b}^k,\rho_b^k,\mat u_{j,b}^k}$, updates $\mat z_i^k$ and $\mat u_i^k$, and evaluates the local stopping metrics $\|\Delta \dof_i^{k+1}\|_\infty$, $\mat r_i^k$, $\mat s_i^k$, and $\mathrm{TOI}_i^k$ (Lines~6--8). These quantities are reported to the controller, which aggregates all worker reports, checks the global stopping criterion, and broadcasts the control signal (Line~9). Worker~$i$ then either terminates if $\sigma=\texttt{end}$ or continues otherwise (Lines~10--12). If the iteration continues, it updates $\rho_b^{k+1}$ for the shared bodies and solves Eq.~(5) for $\dof_i^{k+1}$ (Lines~13--14). After convergence, it assigns the converged shared states and finalizes the frame (Line~15).


\begin{algorithm}[t]\DontPrintSemicolon
\caption{Consensus ADMM on worker $i$}
\label{algorithm:worker_consensus_admm}
\small
\changed{}{
\KwIn{$\mathcal{N}_i$, $\mathcal{B}_{i,\partial}$, $K$}
\nl fetch $\{\mat M_b,\mathrm{id}_b,\dof_b,\dot{\dof}_b\}_{b\in\mathcal{B}_{i,\partial}}$ from $j\in\mathcal{N}_i$\tcp*[r]{init frame}
\nl build local data and $\widetilde{\dof}_i$\;
\nl $\mat z_i^1 \gets \widetilde{\dof}_i, \space \space \mat u_i^1 \gets 0$\tcp*[r]{init ADMM}
\nl \For{$k=1$ \KwTo $K$}{
    \nl \If{$k>1$}{
        \tcp*[l]{fetch shared-body states}
        \nl fetch $\{\dof_{j,b}^k,\rho_b^k,\mat u_{j,b}^k\}_{j\in\mathcal{N}_i,\;b\in\mathcal{B}_{ij}}$ \;
        
        \tcp*[l]{consensus and dual updates}
        \nl update $\mat z_i^k$ and $\mat u_i^k$ using Eq.~\eqref{equ:consensus_solve} and Eq.~\eqref{equ:update_dual}\;
        
        \tcp*[l]{evaluate stopping metrics}
        \nl evaluate $\|\Delta \dof_i^{k+1}\|_\infty$, $\mat r_i^k$, $\mat s_i^k$, and $\mathrm{TOI}_i^k$\;
        
        \tcp*[l]{controller communication}
        \nl report $(\|\Delta \dof_i^{k+1}\|_\infty,\mat r_i^k,\mat s_i^k,\mathrm{TOI}_i^k)$ to controller\;
        \nl receive signal $\sigma$\;
        \nl \If{$\sigma=\texttt{end}$}{
            \nl \textbf{break}
        }
        \nl update $\rho_b^{k+1}$ for $b \in \mathcal B_{i,\partial}$ \tcp*[r]{penalty update} 
    }
    \nl solve Eq.~\eqref{equ:local_solve} for $\dof_i^{k+1}$\tcp*[r]{local solve}
}
\nl assign converged shared states and end the frame\; }
\end{algorithm}


}

\subsection{Control-based Load Balancing} \label{sec:load_balance}
To mitigate load imbalance, we use a lightweight discrete-time PD controller~\cite{1995PID} that updates partition boundaries from per-frame timing feedback.
Each boundary plane $\Gamma_{ij}$ is parameterized by a point $\mathbf{p}_{i,j}$ and a unit normal $\mathbf{n}_{i,j}$ oriented from worker $i$ to worker $j$.
In the two-worker case, let $\tau_i^{k}$ be the smoothed compute time of worker $i$ at frame $k$, and define $\eta^{k}=\tau_1^{k}/\tau_2^{k}$.
We normalize the imbalance as
\begin{equation}
T^{k}=\frac{\eta^{k}-1}{\eta^{k}+1}\in(-1,1),
\end{equation}
where $T^{k}=0$ indicates perfect balance and the sign indicates which worker is slower.
We compute a PD update
\begin{equation}
\Delta p^{k}_{{1,2}} = K_P\,T^{k} + K_D\left(T^{k}-T^{k-1}\right),
\end{equation}
and shift the boundary along its normal,
$\mathbf{p}_{1,2}:=\mathbf{p}_{1,2}+\Delta p^{k}_{1,2}\mathbf{n}_{1,2}$,
to redistribute workload between the two workers.



\section{Experiments}


\begin{table*}[t]
\centering
\caption{Large-scale experiment statistics.
Number of ADMM iterations and timings are averaged over the entire simulation, while collision candidate and contact counts are reported as peak values.
Timing results are reported per frame, where $t_{\text{solve}}$ measures the time spent in local Newton solve, $t_{\text{coll}}$ accounts for collision detection, $t_{\text{sync}}$ denotes synchronization and communication overhead, and $t_{\text{frame}}$ is the total elapsed frame time.
For the rigid-fluid coupling example, the object count reported as $1\mathrm{M}+8$ refers to one million fluid particles interacting with eight rigid bodies.}
\vspace{-6pt}
\resizebox{0.93\textwidth}{!}{
\begin{tabular}{l|cccccccccccc}
\hline
Example               & \# Worker      & \# DoF  & \# Objects  & \#  Triangle          & \# Candidate  & \# Contact  & $h$(s) & \# ADMM Iter & $t_{solve}$(s) & $t_{coll}$(s) & $t_{sync}$(s) & $t_{frame}$(s) \\ \hline
``Pok\'emon''           & 2     & 62.4K   & 5.2K      & 5.7M & 59.8M     & 40.1K      & 0.0125     & 75.85       & 18.40      & 21.90         & 3.41          & 44.06(1.7$\times$)         \\
Food Falling          & 6   & 313.5K    & 26.1K     & 16.1M           & 164.7M       & 222.6K     & 0.01       & 77.57      & 22.29           & 28.88              &  7.03             & 58.64(5.2$\times$)         \\
Drum and Shells       & 2    & 27.9K & 2.3K      & 4.0M             & 53.6M      & 20.0K     & 0.006     & 51.46      &    6.20        &       17.34        &      2.44         & 26.34(1.5$\times$)       \\
Parts and Robotic Arm & 4     &  85.5K   &  7.1K    & 13.0M              & 452.9M       & 338.4K     & 0.01     & 91.52      &   19.69         &   67.77            & 4.66              & 92.76($\infty$)         \\
Rigid-Fluid Coupling  & 4  &  3.1M        & 1.0M + 8    & 219.6K      & 3.3M       & 133.6K     & 0.003      & 10.28      &     0.40       &     0.63          &  0.06             & 1.09         \\ \hline
\end{tabular}
}
\label{tab:large-scale-demo}
\end{table*}


We implemented the distributed simulation system on a multi-node cluster based on a cloud environment. Each node is equipped with an AMD EPYC 7402 CPU (10 cores, 2.79 GHz), 32 GB of system memory and a single NVIDIA RTX 4090 GPU. Nodes are connected via Ethernet with a peak bandwidth of 1 Gb/s. \changed{We conduct experiments to evaluate system scalability and load balancing, as well as the impact of mass-aware penalty parameters on ADMM convergence. Large-scale, contact-rich scenes are used to stress-test penetration-free simulation, and performance is compared against single-node ABD under identical convergence thresholds. The simulation parameters and system settings are detailed in the supplementary material.}{ In this section, we report scaling tests, results on the mass-aware penalty adaptation strategy, and large-scale simulations. Additional results on dynamic load balancing, parameter sensitivity, and partition orientation, as well as detailed experimental settings, are provided in the supplementary material.}

\subsection{Multi-worker Scaling}\label{sec:scaling}
\begin{figure}[t] 
    \includegraphics[width=0.89\columnwidth] {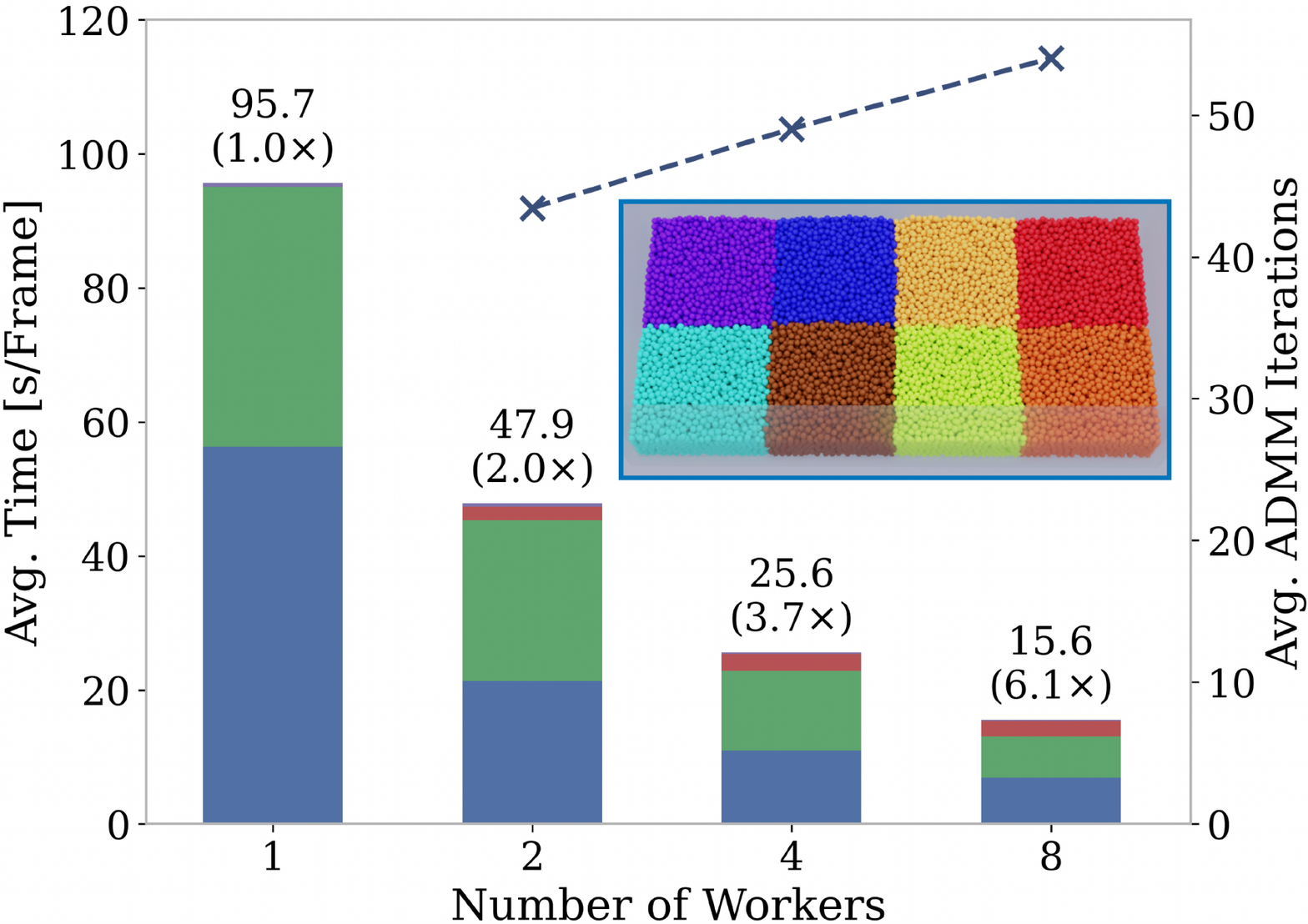} 
    \caption{\changed{}{Scaling results on the same scene partitioned into $\{1,2,4,8\}$ workers. The inset shows the final state of the simulation with 8 workers. The stacked bars report the breakdown of the average time per frame, including Newton solve time (blue), collision detection time (green), synchronization time (red), and other overheads (purple). The dashed line shows the average number of ADMM iterations.}}\label{fig:scaling}
\end{figure}

\changed{We evaluate multi-worker scalability by measuring speedup with increasing workers, considering both weak and strong scaling following standard practice~\cite{wang2020massively, qiu2023sparse}. Benchmark setups are shown in Fig.~\ref{fig:weak-strong-scaling}, with weak scaling maintaining constant per-worker workload and strong scaling fixing the total workload.
}{We evaluate multi-worker scalability by measuring speedup with increasing workers.}
\changed{
For weak scaling, we increase the global problem size proportionally with the number of workers.
A box-shaped container is subdivided into $N$ equal-sized cells, each assigned to one worker, with $1{,}000$ identical spheres initialized above each cell and released under gravity.
We measure the end-to-end runtime over 300 frames for $N\in\{1,2,4,8\}$. 
As shown in Fig.~\ref{fig:scaling}~(left), the total runtime remains nearly constant as the number of workers increases, indicating negligible synchronization overhead and good weak scaling behavior.}{}
We simulate a fixed large-scale scene consisting of $29{,}760$ spheres falling into a square container under gravity.
The number of workers is varied by partitioning the same domain into $N\in\{1,2,4,8\}$. As shown in Fig.~\ref{fig:scaling}, near-linear speedup is achieved with 2 and 4 workers.
When scaling to 8 workers, the speedup decreases to about $6\times$, primarily due to stronger cross-partition coupling under finer partitioning, which leads to more ADMM iterations and a higher proportion of synchronization and interface-handling overhead. \changed{}{For completeness, we further conduct a weak scaling test following \cite{qiu2023sparse, wang2020massively} and report the results in the supplementary material.}

\changed{We evaluate our dynamic load-balancing scheme on a benchmark in which $10{,}780$ star-shaped objects fall into a long, narrow container under gravity, as shown in Fig.~\ref{fig:load_balance_factor}. 
Three four-worker configurations are compared: an uneven partition as a baseline, a uniform partition as a near-optimal static reference, and enabling dynamic load balancing on top of the uneven partition.

Ideal load balance is achieved when all workers have identical computation times, in which case synchronization overhead is minimal and frame time is dominated by computation.
To quantify this effect, we define $\phi$ as the fraction of computation time within the total frame time.
Since objects are initially distributed relatively evenly in this scene, the uniform partition provides a strong static baseline.

As shown in Table~\ref{tab:load-balance}, enabling dynamic load balancing reduces the average frame time by $35\%$ compared to the uneven baseline, and slightly outperforms the uniform partition.
To further characterize load balance during simulation, we define a \emph{balance factor} as the ratio between the average and maximum worker computation times per frame, where values closer to $1$ indicate better balance.
Fig.~\ref{fig:load_balance_factor} shows the evolution of this metric over time for all three configurations.
Starting from the uneven partition, dynamic load balancing rapidly improves the balance factor to the level of the uniform partition and maintains comparable or slightly better balance thereafter, demonstrating both fast response and sustained effectiveness.
}{}

\subsection{Mass-aware Penalty Parameter Adaptation}
\paragraph{Validation under varying mass scales} \changed{We evaluate the mass-aware initialization of the ADMM penalty parameter $\rho$ on a scene consisting of 512 rigid bodies falling and piling into a funnel-shaped container, distributed across two workers.}{This experiment tests the effectiveness of the mass-aware initialization across different mass scales. We use a scene consisting of 512 rigid bodies falling and piling into a funnel-shaped container, distributed across two workers. Fig.~\ref{fig:mass-exp} shows that the mass-aware initialization is robust across all tested mass scales.}

To study robustness across mass scales, we vary body density while keeping the geometry fixed, using five density levels:
$10$, $100$, $1{,}000$, $10{,}000$, and $100{,}000~\mathrm{kg/m^3}$.
For each density level, the initial penalty for body $b$ is set as $\rho_b^{0} = \beta m_b$, where $m_b$ denotes the body mass and $\beta \in \{0.01, 0.1, 1.0, 10, 100\}$.
\changed{We report the average number of ADMM iterations required for convergence in Fig.~\ref{fig:mass-exp}.}{}
\changed{ Across all density levels, $\beta=1.0$ consistently yields the smallest iteration count, indicating that initializing $\rho$ proportional to body mass provides a robust baseline over a wide range of mass magnitudes.
We also observe slower convergence in low-density settings.
With smaller masses, the ABD system becomes more ill-conditioned under the same scene configuration and time step, which increases the difficulty of the local Newton solves and leads to a higher number of ADMM iterations.}{In particular, $\beta=1.0$ gives the lowest or near-lowest iteration count across all density levels, while $\beta=0.1$ also performs well. This suggests that the mass-aware initialization is robust across a wide range of mass scales. The main outlier is the lightest case, $10~\mathrm{kg/m^3}$, with $\beta=10^{-2}$. In this case, $\rho_b^0$ becomes too small, so the consensus penalty is too weak to effectively enforce agreement among workers, which leads to slower ADMM convergence.}

\paragraph{Ablation of residual-driven $\rho$ adaptation.}
We evaluate  residual-driven adaptation of the ADMM penalty parameter in the same two-worker scene with 512 rigid bodies.
The mass-based initialization is fixed to $\beta=1.0$ (i.e., $\rho_b^{0}=m_b$), and we compare convergence with and without residual-driven adaptation.
For each density level, the average number of ADMM iterations required to reach the same stopping criteria is measured.
As shown in Fig.~\ref{fig:rho_adap}, enabling residual-driven adaptation reduces the iteration count by $22\%$–$80\%$ across all density levels compared to using a fixed penalty parameter.

\changed{}{
\paragraph{Validation under material heterogeneity} 
Our method remains robust under large mass and stiffness contrasts. We construct a two-worker scene with 200 bodies, divided into three groups with $(m,E)=(1,10^8)$, $(100,10^{10})$, and $(10^4,10^{12})$. The scene is visualized in Fig.~\ref{fig:exp-vary-mass}. We compare this setting with a homogeneous baseline where all bodies use $(100,10^{10})$. In the heterogeneous case, the distributed solver requires about $2\times$ more ADMM iterations and $2.5\times$ more Newton iterations as reported in Table~\ref{tab:vary-mass-young}. This increase mainly reflects the greater difficulty of the underlying ABD problem, rather than the distributed ADMM scheme itself. To verify this, we also run the same setting on a single worker, where the Newton iteration count likewise increases by about $2.5\times$. We further vary the mass-aware coefficient $\beta$ to evaluate the proposed initialization. Detailed results are provided in the supplementary material.}


\begin{table}[t]
\centering
\caption{\changed{}{Effect of mass and stiffness heterogeneity on convergence. Here, 2W and 1W denote two-worker and single-worker settings, respectively; ``hetero". denotes the heterogeneous setting, and ``uniform" denotes the uniform setting.} }
\label{tab:vary-mass-young}
\setlength{\tabcolsep}{4pt}
\changed{}{
\begin{tabular}{lcc}
\toprule
Case & \# ADMM Iter & \# Newton Iter \\
\midrule
2W, hetero.  & 66.82 & 147.49 \\
2W, uniform  & 36.11 & 60.13 \\
1W, hetero.  & --    & 94.72 \\
1W, uniform  & --    & 37.15 \\
\bottomrule
\end{tabular} }

\end{table}


\subsection{Large-scale Distributed Simulation} 

Large-scale ABD simulation scenarios are used to demonstrate the capability and robustness of the proposed system, including numerical stability, high collision resolution, and effective scaling in distributed environments. Table~\ref{tab:large-scale-demo} summarizes detailed statistics and configuration parameters for all experiments. \changed{}{To clarify the reported collision statistics, we refer to broad-phase outputs as candidate collision pairs and to pairs retained after narrow-phase filtering as active contact pairs~\cite{ericson2004real}.}

\paragraph{``Pok\'emon''.}
We simulate a ``Pok\'emon'' scenario in which objects fall through a narrow funnel and interact with two perforated trays inside a container, as illustrated in Fig.~\ref{fig:teaser}.
The simulation involves $5{,}203$ rigid bodies with $5.7$M triangles, distributed across two workers.
Bodies are released from above, pass through the funnel, and land on two stacked rotating perforated plates.
Some bodies remain on the plates, while others fall through the holes and reach the container bottom.
During the simulation, we handle up to $59.8$M candidate collision pairs and $40.1$K active contact pairs. Interpenetration does not occur at any point during the simulation. Average wall-clock time is 44.06s per frame for this example, a 1.7$\times$ improvement over single-worker execution.

\begin{figure*}[p] 
  \centering
  \includegraphics[width=\textwidth,height=1.2\textheight,keepaspectratio]{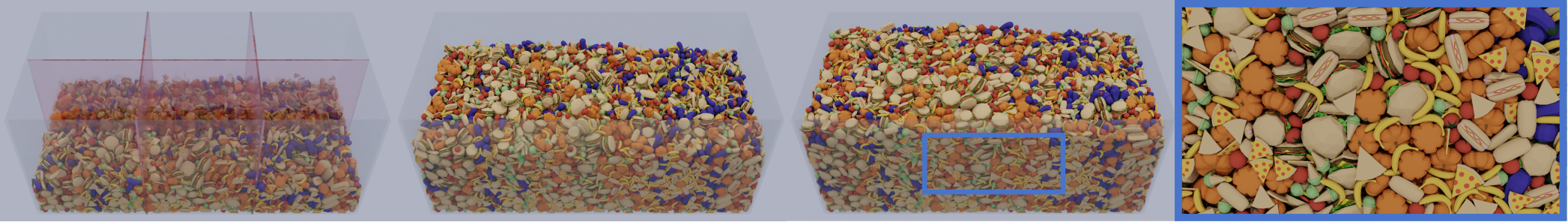}
  \caption{\emph{Falling foods.} A falling and piling scene with food-shaped rigid bodies, containing over $16M$ triangles and reaching a peak of 164M collision candidates and $222K$ active contact pairs. Using six workers, the distributed solver achieves a $5.2\times$ speedup over a single-worker run. Even under such dense contact interactions, our system remains globally penetration-free and continues to exhibit near-linear scaling across workers. Note that the red planes shown in the leftmost sub-figure indicate the spatial partitioning. The same convention is used in Figs. ~\ref{fig:demo_arm} and ~\ref{fig:demo_shells}.
}
  \label{fig:demo_food}
\end{figure*}

\paragraph{Falling foods.}
We simulate a large-scale dropping scenario consisting of $27{,}163$ ABD bodies distributed across six workers, leading to extreme geometric complexity and collision density as the container fills. Fig.~\ref{fig:demo_food} shows the resulting dynamics. The simulation involves $16$M triangles and reaches up to $164$M collision candidates with $222$K active contact pairs. Despite the heavy collision workload, penetration-free configurations are maintained throughout the simulation. In this simulation, the average frame time is 58.64 s, achieving a 5.2$\times$ speedup over a single-worker baseline.

\begin{figure*}[p] 
  \centering
  \includegraphics[width=\textwidth,height=0.95\textheight,keepaspectratio]{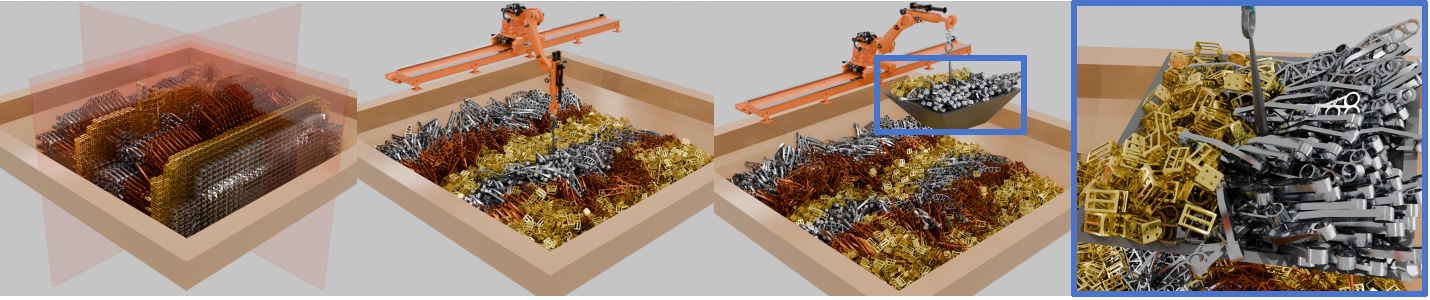}
  \caption{\emph{Industrial parts and robotic arm}. A large-scale scenario with $7K$ industrial parts and a robotic arm, totaling over $13M$ triangles and hundreds of millions of contact candidates. The scene cannot be executed on a single machine due to memory constraints. By distributing geometry and contact processing across workers, our system significantly reduces per-worker memory footprint, enabling simulation at scales that exceed the capacity of a single machine.}
  \label{fig:demo_arm}
\end{figure*}

\paragraph{Industrial parts and robotic arm.} We also test an industrial part lifting task driven by a robotic arm, simulated on four workers with $7{,}124$ ABD bodies.
A container with a pull ring is placed at the scene center, and the robotic arm successfully grasps the ring and lifts the container.
The simulation includes over $13$M triangles and produces $452$M collision candidates and $338$K active contact pairs during the dropping and lifting stages. Fig.~\ref{fig:demo_arm} contains rendered snapshots from this experiment. On average, this case takes 92.76s per frame on four workers, whereas a single-GPU setup cannot accommodate the case due to insufficient GPU memory. The simulation remains penetration-free across all frames.

\begin{figure*}[p] 
  \centering
  \includegraphics[width=\textwidth,height=0.95\textheight,keepaspectratio]{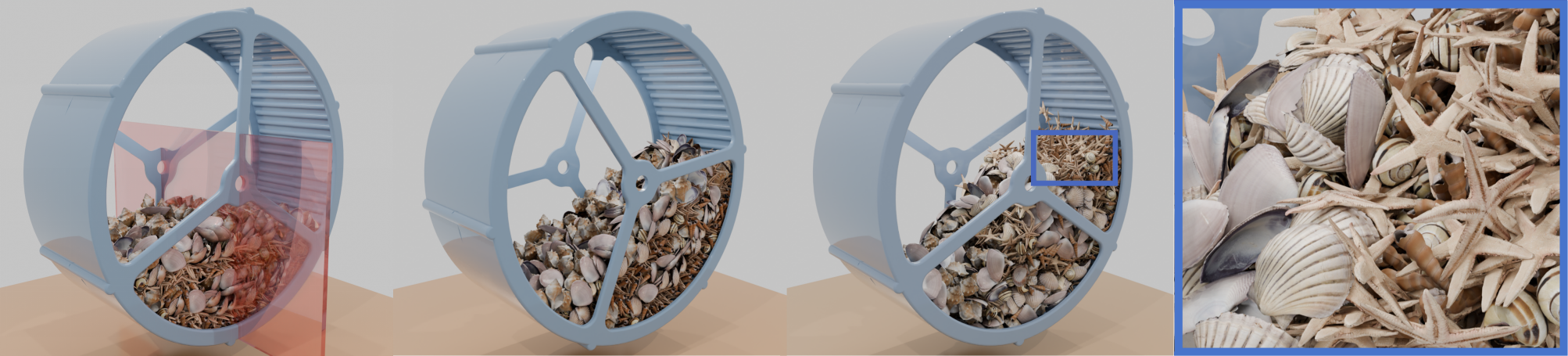}
  \caption{\emph{Rotating drum with shells}. Dense-contact simulation of $2K$ complex-shaped rigid bodies(e.g., starfish, conches, and shells) continuously agitated in a rotating drum, generating up to $54M$ collision candidates. The two-worker distributed solver achieves a $1.5\times$ speedup over a single-machine run. Even in this challenging setting with thin shells and sharp features undergoing sustained agitation, our solver maintains penetration-free contact throughout the simulation.}
  \label{fig:demo_shells}
\end{figure*}

\paragraph{Rotating drum with shells.} We simulate a rotating-drum scenario on two workers with $2{,}323$ rigid bodies and $4$M triangles. As in Fig.~\ref{fig:demo_shells}, the drum acts as a moving boundary and continuously agitates bodies with sharp, non-convex shapes, causing repeated mixing and piling inside the drum.
The complex body geometries sustain a heavy collision workload and generate up to $54$M collision candidates.
Collision detection dominates the computation and accounts for $66\%$ of the total runtime, as reported in Table~\ref{tab:large-scale-demo}. With two workers, the simulation remains penetration-free and runs at 26.34s per frame on average, and the resulting speedup over a single worker is 1.5 $\times$. 

\paragraph{Rigid-fluid coupling.} We further evaluate the scalability of our distributed ABD framework by applying it to fluid simulation and rigid-fluid coupling.
We implement an IISPH~\cite{ihmsen2013implicit} solver and integrate it into our distributed system according to ~\cite{rustico2012advances}. As for the rigid-fluid coupling, we implement a simple impulse-based method~\cite{becker2009direct} to compute the two-way coupling force at the beginning of each frame and apply the resulting coupling force to both fluid particles and rigid bodies as external forces. In this case, we use $h=0.003$ and $\theta=0.0008$ and assign 4 workers with $1$M fluid particles and $220$K triangles for ABD objects. The results are visualized in Fig.~\ref{fig:demo_coupling}. The average run time of this experiment is 1.09s per frame.



\section{Conclusion}

In this work, we present a distributed ABD framework based on ADMM that enables penetration-free simulation across multiple GPU workers, allowing large-scale contact handling in a distributed environment.
To address the sensitivity of ADMM to the penalty parameter, we initialize $\rho$ based on object mass and further refine it through residual-driven adaptation, significantly accelerating convergence.
With a lightweight load-balancing scheme, experiments on large-scale, contact-rich scenarios demonstrate stable, penetration-free simulation and near-linear speedups in multi-node distributed settings.

\paragraph{Limitations and future work}
Our current implementation relies on blocking communication and does not overlap data transfer with local computation, causing communication latency to directly contribute to the overall runtime. 
Incorporating asynchronous communication together with latency-hiding mechanisms could substantially reduce the effective communication overhead and further improve scalability.
Moreover, our experiments indicate that finer partitioning leads to an increased number of ADMM iterations, potentially offsetting the benefits of improved load balance. Exploring more effective warm-start strategies or partition-aware initialization will be a valuable direction.
Finally, ADMM may converge slowly when high accuracy is required, often exhibiting a long tail near convergence. To this end, we plan to explore Anderson acceleration and related techniques to mitigate this long-tail slowdown and reduce the total number of iterations.


\begin{acks}

We would like to sincerely thank all the reviewers for their valuable and constructive comments and suggestions. Weiwei Xu is partially supported by NSFC (92570206, 62421003), and State Key Laboratory of CAD\&CG, Zhejiang University. Yin Yang is partially supported by NSF 2301040. Yifan Peng is partially supported by the National Natural Science Foundation of China (62322217) and the Innovation and Technology Fund of Hong Kong (MHP/313/24).
\end{acks}


\bibliographystyle{ACM-Reference-Format}





\begin{thebibliography}{58}


\ifx \showCODEN    \undefined \def \showCODEN     #1{\unskip}     \fi
\ifx \showISBNx    \undefined \def \showISBNx     #1{\unskip}     \fi
\ifx \showISBNxiii \undefined \def \showISBNxiii  #1{\unskip}     \fi
\ifx \showISSN     \undefined \def \showISSN      #1{\unskip}     \fi
\ifx \showLCCN     \undefined \def \showLCCN      #1{\unskip}     \fi
\ifx \shownote     \undefined \def \shownote      #1{#1}          \fi
\ifx \showarticletitle \undefined \def \showarticletitle #1{#1}   \fi
\ifx \showURL      \undefined \def \showURL       {\relax}        \fi
\providecommand\bibfield[2]{#2}
\providecommand\bibinfo[2]{#2}
\providecommand\natexlab[1]{#1}
\providecommand\showeprint[2][]{arXiv:#2}

\bibitem[Barreiros et~al\mbox{.}(2025)]%
        {barreiros2025learning}
\bibfield{author}{\bibinfo{person}{Jose~A Barreiros},
  \bibinfo{person}{Aykut~{\"O}zg{\"u}n {\"O}nol}, \bibinfo{person}{Mengchao
  Zhang}, \bibinfo{person}{Sam Creasey}, \bibinfo{person}{Aimee Goncalves},
  \bibinfo{person}{Andrew Beaulieu}, \bibinfo{person}{Aditya Bhat},
  \bibinfo{person}{Kate~M Tsui}, {and} \bibinfo{person}{Alex Alspach}.}
  \bibinfo{year}{2025}\natexlab{}.
\newblock \showarticletitle{Learning contact-rich whole-body manipulation with
  example-guided reinforcement learning}.
\newblock \bibinfo{journal}{\emph{Science Robotics}} \bibinfo{volume}{10},
  \bibinfo{number}{105} (\bibinfo{year}{2025}), \bibinfo{pages}{eads6790}.
\newblock


\bibitem[Becker et~al\mbox{.}(2009)]%
        {becker2009direct}
\bibfield{author}{\bibinfo{person}{Markus Becker}, \bibinfo{person}{Hendrik
  Tessendorf}, {and} \bibinfo{person}{Matthias Teschner}.}
  \bibinfo{year}{2009}\natexlab{}.
\newblock \showarticletitle{Direct forcing for lagrangian rigid-fluid
  coupling}.
\newblock \bibinfo{journal}{\emph{IEEE Transactions on Visualization and
  Computer Graphics}} \bibinfo{volume}{15}, \bibinfo{number}{3}
  (\bibinfo{year}{2009}), \bibinfo{pages}{493--503}.
\newblock


\bibitem[Boyd et~al\mbox{.}(2011)]%
        {boyd2011distributed}
\bibfield{author}{\bibinfo{person}{Stephen Boyd}, \bibinfo{person}{Neal
  Parikh}, \bibinfo{person}{Eric Chu}, \bibinfo{person}{Borja Peleato},
  \bibinfo{person}{Jonathan Eckstein}, {et~al\mbox{.}}}
  \bibinfo{year}{2011}\natexlab{}.
\newblock \showarticletitle{Distributed optimization and statistical learning
  via the alternating direction method of multipliers}.
\newblock \bibinfo{journal}{\emph{Foundations and Trends{\textregistered} in
  Machine learning}} \bibinfo{volume}{3}, \bibinfo{number}{1}
  (\bibinfo{year}{2011}), \bibinfo{pages}{1--122}.
\newblock


\bibitem[Brown(2020)]%
        {brown2020distributed}
\bibfield{author}{\bibinfo{person}{Alexander James~Ronald Brown}.}
  \bibinfo{year}{2020}\natexlab{}.
\newblock \emph{\bibinfo{title}{Distributed real-time physics for scalable and
  streamed games and simulation}}.
\newblock \bibinfo{thesistype}{Ph.\,D. Dissertation}.
  \bibinfo{school}{Newcastle University}.
\newblock


\bibitem[Brown and Narain(2021)]%
        {brown2021wrapd}
\bibfield{author}{\bibinfo{person}{George~E Brown} {and} \bibinfo{person}{Rahul
  Narain}.} \bibinfo{year}{2021}\natexlab{}.
\newblock \showarticletitle{Wrapd: weighted rotation-aware admm for
  parameterization and deformation}.
\newblock \bibinfo{journal}{\emph{ACM Transactions on Graphics (TOG)}}
  \bibinfo{volume}{40}, \bibinfo{number}{4} (\bibinfo{year}{2021}),
  \bibinfo{pages}{1--14}.
\newblock


\bibitem[Chan and Mathew(1994)]%
        {chan1994domain}
\bibfield{author}{\bibinfo{person}{Tony~F Chan} {and} \bibinfo{person}{Tarek~P
  Mathew}.} \bibinfo{year}{1994}\natexlab{}.
\newblock \showarticletitle{Domain decomposition algorithms}.
\newblock \bibinfo{journal}{\emph{Acta numerica}}  \bibinfo{volume}{3}
  (\bibinfo{year}{1994}), \bibinfo{pages}{61--143}.
\newblock


\bibitem[Chen et~al\mbox{.}(2025)]%
        {chen2025offset}
\bibfield{author}{\bibinfo{person}{Anka~He Chen}, \bibinfo{person}{Jerry Hsu},
  \bibinfo{person}{Ziheng Liu}, \bibinfo{person}{Miles Macklin},
  \bibinfo{person}{Yin Yang}, {and} \bibinfo{person}{Cem Yuksel}.}
  \bibinfo{year}{2025}\natexlab{}.
\newblock \showarticletitle{Offset Geometric Contact}.
\newblock \bibinfo{journal}{\emph{ACM Transactions on Graphics (TOG)}}
  \bibinfo{volume}{44}, \bibinfo{number}{4} (\bibinfo{year}{2025}),
  \bibinfo{pages}{1--21}.
\newblock


\bibitem[Chen et~al\mbox{.}(2022)]%
        {chen2022unified}
\bibfield{author}{\bibinfo{person}{Yunuo Chen}, \bibinfo{person}{Minchen Li},
  \bibinfo{person}{Lei Lan}, \bibinfo{person}{Hao Su}, \bibinfo{person}{Yin
  Yang}, {and} \bibinfo{person}{Chenfanfu Jiang}.}
  \bibinfo{year}{2022}\natexlab{}.
\newblock \showarticletitle{A unified newton barrier method for multibody
  dynamics}.
\newblock \bibinfo{journal}{\emph{ACM Transactions on Graphics (TOG)}}
  \bibinfo{volume}{41}, \bibinfo{number}{4} (\bibinfo{year}{2022}),
  \bibinfo{pages}{1--14}.
\newblock


\bibitem[Chen et~al\mbox{.}(2023)]%
        {FrictionShockChen23}
\bibfield{author}{\bibinfo{person}{Yi-Lu Chen}, \bibinfo{person}{Micka\"{e}l
  Ly}, {and} \bibinfo{person}{Chris Wojtan}.} \bibinfo{year}{2023}\natexlab{}.
\newblock \showarticletitle{Unified treatment of contact, friction and
  shock-propagation in rigid body animation} \emph{(\bibinfo{series}{SCA
  '23})}. \bibinfo{publisher}{Association for Computing Machinery},
  \bibinfo{address}{New York, NY, USA}, Article \bibinfo{articleno}{5},
  \bibinfo{numpages}{2}~pages.
\newblock
\showISBNx{9798400702686}
\href{https://doi.org/10.1145/3606037.3606836}{doi:\nolinkurl{10.1145/3606037.3606836}}


\bibitem[Daviet(2020)]%
        {daviet2020simple}
\bibfield{author}{\bibinfo{person}{Gilles Daviet}.}
  \bibinfo{year}{2020}\natexlab{}.
\newblock \showarticletitle{Simple and scalable frictional contacts for thin
  nodal objects}.
\newblock \bibinfo{journal}{\emph{ACM Transactions on Graphics (TOG)}}
  \bibinfo{volume}{39}, \bibinfo{number}{4} (\bibinfo{year}{2020}),
  \bibinfo{pages}{61--1}.
\newblock


\bibitem[Daviet(2023)]%
        {daviet2023interactive}
\bibfield{author}{\bibinfo{person}{Gilles Daviet}.}
  \bibinfo{year}{2023}\natexlab{}.
\newblock \showarticletitle{Interactive hair simulation on the GPU using ADMM}.
  In \bibinfo{booktitle}{\emph{ACM SIGGRAPH 2023 Conference Proceedings}}.
  \bibinfo{pages}{1--11}.
\newblock


\bibitem[Dean and Ghemawat(2008)]%
        {dean2008mapreduce}
\bibfield{author}{\bibinfo{person}{Jeffrey Dean} {and} \bibinfo{person}{Sanjay
  Ghemawat}.} \bibinfo{year}{2008}\natexlab{}.
\newblock \showarticletitle{MapReduce: simplified data processing on large
  clusters}.
\newblock \bibinfo{journal}{\emph{Commun. ACM}} \bibinfo{volume}{51},
  \bibinfo{number}{1} (\bibinfo{year}{2008}), \bibinfo{pages}{107--113}.
\newblock


\bibitem[Ericson(2004)]%
        {ericson2004real}
\bibfield{author}{\bibinfo{person}{Christer Ericson}.}
  \bibinfo{year}{2004}\natexlab{}.
\newblock \bibinfo{booktitle}{\emph{Real-time collision detection}}.
\newblock \bibinfo{publisher}{Crc Press}.
\newblock


\bibitem[Fang et~al\mbox{.}(2019)]%
        {fang2019silly}
\bibfield{author}{\bibinfo{person}{Yu Fang}, \bibinfo{person}{Minchen Li},
  \bibinfo{person}{Ming Gao}, {and} \bibinfo{person}{Chenfanfu Jiang}.}
  \bibinfo{year}{2019}\natexlab{}.
\newblock \showarticletitle{Silly rubber: an implicit material point method for
  simulating non-equilibrated viscoelastic and elastoplastic solids}.
\newblock \bibinfo{journal}{\emph{ACM Transactions on Graphics (TOG)}}
  \bibinfo{volume}{38}, \bibinfo{number}{4} (\bibinfo{year}{2019}),
  \bibinfo{pages}{1--13}.
\newblock


\bibitem[Ferguson et~al\mbox{.}(2021)]%
        {ferguson2021intersection}
\bibfield{author}{\bibinfo{person}{Zachary Ferguson}, \bibinfo{person}{Minchen
  Li}, \bibinfo{person}{Teseo Schneider}, \bibinfo{person}{Francisca
  Gil-Ureta}, \bibinfo{person}{Timothy Langlois}, \bibinfo{person}{Chenfanfu
  Jiang}, \bibinfo{person}{Denis Zorin}, \bibinfo{person}{Danny~M Kaufman},
  {and} \bibinfo{person}{Daniele Panozzo}.} \bibinfo{year}{2021}\natexlab{}.
\newblock \showarticletitle{Intersection-free rigid body dynamics}.
\newblock \bibinfo{journal}{\emph{ACM Transactions on Graphics}}
  \bibinfo{volume}{40}, \bibinfo{number}{4} (\bibinfo{year}{2021}).
\newblock


\bibitem[Fetterly et~al\mbox{.}(2009)]%
        {fetterly2009dryadlinq}
\bibfield{author}{\bibinfo{person}{Yuan Yu Michael Isard~Dennis Fetterly},
  \bibinfo{person}{Mihai Budiu}, \bibinfo{person}{{\'U}lfar Erlingsson}, {and}
  \bibinfo{person}{Pradeep Kumar Gunda~Jon Currey}.}
  \bibinfo{year}{2009}\natexlab{}.
\newblock \showarticletitle{DryadLINQ: A system for general-purpose distributed
  data-parallel computing using a high-level language}.
\newblock \bibinfo{journal}{\emph{Proc. LSDS-IR}}  \bibinfo{volume}{8}
  (\bibinfo{year}{2009}), \bibinfo{pages}{1--14}.
\newblock


\bibitem[Gropp(2001)]%
        {gropp2001mpi}
\bibfield{author}{\bibinfo{person}{William~D. Gropp}.}
  \bibinfo{year}{2001}\natexlab{}.
\newblock \showarticletitle{Learning from the Success of {MPI}}. In
  \bibinfo{booktitle}{\emph{High Performance Computing --- HiPC 2001}}
  \emph{(\bibinfo{series}{Lecture Notes in Computer Science},
  Vol.~\bibinfo{volume}{2228})}, \bibfield{editor}{\bibinfo{person}{Burkhard
  Monien}, \bibinfo{person}{Viktor~K. Prasanna}, {and} \bibinfo{person}{Sriram
  Vajapeyam}} (Eds.). \bibinfo{publisher}{Springer}, \bibinfo{pages}{81--94}.
\newblock
\href{https://doi.org/10.1007/3-540-45307-5_8}{doi:\nolinkurl{10.1007/3-540-45307-5_8}}


\bibitem[Guo et~al\mbox{.}(2024)]%
        {guo2024barrier}
\bibfield{author}{\bibinfo{person}{Dewen Guo}, \bibinfo{person}{Minchen Li},
  \bibinfo{person}{Yin Yang}, \bibinfo{person}{Sheng Li}, {and}
  \bibinfo{person}{Guoping Wang}.} \bibinfo{year}{2024}\natexlab{}.
\newblock \showarticletitle{Barrier-Augmented Lagrangian for GPU-based
  Elastodynamic Contact}.
\newblock \bibinfo{journal}{\emph{ACM Transactions on Graphics (TOG)}}
  \bibinfo{volume}{43}, \bibinfo{number}{6} (\bibinfo{year}{2024}),
  \bibinfo{pages}{1--17}.
\newblock


\bibitem[Huang et~al\mbox{.}(2024a)]%
        {huang2024gipc}
\bibfield{author}{\bibinfo{person}{Kemeng Huang}, \bibinfo{person}{Floyd~M
  Chitalu}, \bibinfo{person}{Huancheng Lin}, {and} \bibinfo{person}{Taku
  Komura}.} \bibinfo{year}{2024}\natexlab{a}.
\newblock \showarticletitle{GIPC: Fast and stable gauss-newton optimization of
  IPC barrier energy}.
\newblock \bibinfo{journal}{\emph{ACM Transactions on Graphics}}
  \bibinfo{volume}{43}, \bibinfo{number}{2} (\bibinfo{year}{2024}),
  \bibinfo{pages}{1--18}.
\newblock


\bibitem[Huang et~al\mbox{.}(2024b)]%
        {huang2024stiffgipc}
\bibfield{author}{\bibinfo{person}{Kemeng Huang}, \bibinfo{person}{Xinyu Lu},
  \bibinfo{person}{Huancheng Lin}, \bibinfo{person}{Taku Komura}, {and}
  \bibinfo{person}{Minchen Li}.} \bibinfo{year}{2024}\natexlab{b}.
\newblock \showarticletitle{StiffGIPC: Advancing GPU IPC for stiff
  affine-deformable simulation}.
\newblock \bibinfo{journal}{\emph{arXiv preprint arXiv:2411.06224}}
  (\bibinfo{year}{2024}).
\newblock


\bibitem[Ihmsen et~al\mbox{.}(2013)]%
        {ihmsen2013implicit}
\bibfield{author}{\bibinfo{person}{Markus Ihmsen}, \bibinfo{person}{Jens
  Cornelis}, \bibinfo{person}{Barbara Solenthaler},
  \bibinfo{person}{Christopher Horvath}, {and} \bibinfo{person}{Matthias
  Teschner}.} \bibinfo{year}{2013}\natexlab{}.
\newblock \showarticletitle{Implicit incompressible SPH}.
\newblock \bibinfo{journal}{\emph{IEEE transactions on visualization and
  computer graphics}} \bibinfo{volume}{20}, \bibinfo{number}{3}
  (\bibinfo{year}{2013}), \bibinfo{pages}{426--435}.
\newblock


\bibitem[Inglis et~al\mbox{.}(2017)]%
        {inglis2017primal}
\bibfield{author}{\bibinfo{person}{Tiffany Inglis}, \bibinfo{person}{M-L
  Eckert}, \bibinfo{person}{James Gregson}, {and} \bibinfo{person}{Nils
  Thuerey}.} \bibinfo{year}{2017}\natexlab{}.
\newblock \showarticletitle{Primal-dual optimization for fluids}. In
  \bibinfo{booktitle}{\emph{Computer Graphics Forum}},
  Vol.~\bibinfo{volume}{36}. Wiley Online Library, \bibinfo{pages}{354--368}.
\newblock


\bibitem[Ji et~al\mbox{.}(2025)]%
        {ji2025gpu}
\bibfield{author}{\bibinfo{person}{Harim Ji}, \bibinfo{person}{Hyunsu Kim},
  \bibinfo{person}{Jeongmin Lee}, \bibinfo{person}{Somang Lee},
  \bibinfo{person}{Seoki An}, \bibinfo{person}{Jinuk Heo},
  \bibinfo{person}{Youngseon Lee}, \bibinfo{person}{Yongseok Lee}, {and}
  \bibinfo{person}{Dongjun Lee}.} \bibinfo{year}{2025}\natexlab{}.
\newblock \showarticletitle{GPU-Accelerated Subsystem-Based ADMM for
  Large-Scale Interactive Simulation}. In \bibinfo{booktitle}{\emph{ICRA 2025
  Workshop''Handy Moves: Dexterity in Multi-Fingered Hands''Paper Submission}}.
\newblock


\bibitem[Kale(2023)]%
        {kale2023distributed}
\bibfield{author}{\bibinfo{person}{Manas~Anand Kale}.}
  \bibinfo{year}{2023}\natexlab{}.
\newblock \bibinfo{booktitle}{\emph{Distributed Simulation of Large Multi-body
  Systems}}.
\newblock \bibinfo{publisher}{McGill University (Canada)}.
\newblock


\bibitem[Lan et~al\mbox{.}(2022a)]%
        {lan2022affine}
\bibfield{author}{\bibinfo{person}{Lei Lan}, \bibinfo{person}{Danny~M.
  Kaufman}, \bibinfo{person}{Minchen Li}, \bibinfo{person}{Chenfanfu Jiang},
  {and} \bibinfo{person}{Yin Yang}.} \bibinfo{year}{2022}\natexlab{a}.
\newblock \showarticletitle{Affine body dynamics: fast, stable and
  intersection-free simulation of stiff materials}.
\newblock  \bibinfo{volume}{41}, \bibinfo{number}{4}, Article
  \bibinfo{articleno}{67} (\bibinfo{date}{July} \bibinfo{year}{2022}),
  \bibinfo{numpages}{14}~pages.
\newblock
\showISSN{0730-0301}
\href{https://doi.org/10.1145/3528223.3530064}{doi:\nolinkurl{10.1145/3528223.3530064}}


\bibitem[Lan et~al\mbox{.}(2023)]%
        {lan2023second}
\bibfield{author}{\bibinfo{person}{Lei Lan}, \bibinfo{person}{Minchen Li},
  \bibinfo{person}{Chenfanfu Jiang}, \bibinfo{person}{Huamin Wang}, {and}
  \bibinfo{person}{Yin Yang}.} \bibinfo{year}{2023}\natexlab{}.
\newblock \showarticletitle{Second-order Stencil Descent for Interior-point
  Hyperelasticity}.
\newblock \bibinfo{journal}{\emph{ACM Trans. Graph.}} \bibinfo{volume}{42},
  \bibinfo{number}{4}, Article \bibinfo{articleno}{108} (\bibinfo{date}{July}
  \bibinfo{year}{2023}), \bibinfo{numpages}{16}~pages.
\newblock
\showISSN{0730-0301}
\href{https://doi.org/10.1145/3592104}{doi:\nolinkurl{10.1145/3592104}}


\bibitem[Lan et~al\mbox{.}(2024)]%
        {lan2024efficient}
\bibfield{author}{\bibinfo{person}{Lei Lan}, \bibinfo{person}{Zixuan Lu},
  \bibinfo{person}{Jingyi Long}, \bibinfo{person}{Chun Yuan},
  \bibinfo{person}{Xuan Li}, \bibinfo{person}{Xiaowei He},
  \bibinfo{person}{Huamin Wang}, \bibinfo{person}{Chenfanfu Jiang}, {and}
  \bibinfo{person}{Yin Yang}.} \bibinfo{year}{2024}\natexlab{}.
\newblock \showarticletitle{Efficient GPU cloth simulation with non-distance
  barriers and subspace reuse}.
\newblock \bibinfo{journal}{\emph{arXiv preprint arXiv:2403.19272}}
  (\bibinfo{year}{2024}).
\newblock


\bibitem[Lan et~al\mbox{.}(2022b)]%
        {lan2022penetration}
\bibfield{author}{\bibinfo{person}{Lei Lan}, \bibinfo{person}{Guanqun Ma},
  \bibinfo{person}{Yin Yang}, \bibinfo{person}{Changxi Zheng},
  \bibinfo{person}{Minchen Li}, {and} \bibinfo{person}{Chenfanfu Jiang}.}
  \bibinfo{year}{2022}\natexlab{b}.
\newblock \showarticletitle{Penetration-free projective dynamics on the GPU}.
\newblock \bibinfo{journal}{\emph{ACM Transactions on Graphics (TOG)}}
  \bibinfo{volume}{41}, \bibinfo{number}{4} (\bibinfo{year}{2022}),
  \bibinfo{pages}{1--16}.
\newblock


\bibitem[Lee et~al\mbox{.}(2023)]%
        {lee2023modular}
\bibfield{author}{\bibinfo{person}{Jeongmin Lee}, \bibinfo{person}{Minji Lee},
  {and} \bibinfo{person}{Dongjun Lee}.} \bibinfo{year}{2023}\natexlab{}.
\newblock \showarticletitle{Modular and parallelizable multibody physics
  simulation via subsystem-based ADMM}.
\newblock \bibinfo{journal}{\emph{arXiv preprint arXiv:2302.14344}}
  (\bibinfo{year}{2023}).
\newblock


\bibitem[Lesser et~al\mbox{.}(2022)]%
        {lesser2022loki}
\bibfield{author}{\bibinfo{person}{Steve Lesser}, \bibinfo{person}{Alexey
  Stomakhin}, \bibinfo{person}{Gilles Daviet}, \bibinfo{person}{Joel Wretborn},
  \bibinfo{person}{John Edholm}, \bibinfo{person}{Noh-Hoon Lee},
  \bibinfo{person}{Eston Schweickart}, \bibinfo{person}{Xiao Zhai},
  \bibinfo{person}{Sean Flynn}, {and} \bibinfo{person}{Andrew Moffat}.}
  \bibinfo{year}{2022}\natexlab{}.
\newblock \showarticletitle{Loki: a unified multiphysics simulation framework
  for production}.
\newblock \bibinfo{journal}{\emph{ACM Transactions on Graphics (TOG)}}
  \bibinfo{volume}{41}, \bibinfo{number}{4} (\bibinfo{year}{2022}),
  \bibinfo{pages}{1--20}.
\newblock


\bibitem[Li et~al\mbox{.}(2020c)]%
        {pcloth20}
\bibfield{author}{\bibinfo{person}{Cheng Li}, \bibinfo{person}{Min Tang},
  \bibinfo{person}{Ruofeng Tong}, \bibinfo{person}{Ming Cai},
  \bibinfo{person}{Jieyi Zhao}, {and} \bibinfo{person}{Dinesh Manocha}.}
  \bibinfo{year}{2020}\natexlab{c}.
\newblock \showarticletitle{{P-Cloth}: Interactive Cloth Simulation on
  Multi-{GPU} Systems using Dynamic Matrix Assembly and Pipelined Implicit
  Integrators}.
\newblock \bibinfo{journal}{\emph{ACM Transaction on Graphics (Proceedings of
  SIGGRAPH Asia)}} \bibinfo{volume}{39}, \bibinfo{number}{6}
  (\bibinfo{date}{December} \bibinfo{year}{2020}), \bibinfo{pages}{180:1--15}.
\newblock


\bibitem[Li et~al\mbox{.}(2020a)]%
        {Li2020IPC}
\bibfield{author}{\bibinfo{person}{Minchen Li}, \bibinfo{person}{Zachary
  Ferguson}, \bibinfo{person}{Teseo Schneider}, \bibinfo{person}{Timothy
  Langlois}, \bibinfo{person}{Denis Zorin}, \bibinfo{person}{Daniele Panozzo},
  \bibinfo{person}{Chenfanfu Jiang}, {and} \bibinfo{person}{Danny~M. Kaufman}.}
  \bibinfo{year}{2020}\natexlab{a}.
\newblock \showarticletitle{Incremental Potential Contact: Intersection- and
  Inversion-free Large Deformation Dynamics}.
\newblock \bibinfo{journal}{\emph{ACM Trans. Graph. (SIGGRAPH)}}
  \bibinfo{volume}{39}, \bibinfo{number}{4}, Article \bibinfo{articleno}{49}
  (\bibinfo{year}{2020}).
\newblock


\bibitem[Li et~al\mbox{.}(2020b)]%
        {li2020codimensional}
\bibfield{author}{\bibinfo{person}{Minchen Li}, \bibinfo{person}{Danny~M
  Kaufman}, {and} \bibinfo{person}{Chenfanfu Jiang}.}
  \bibinfo{year}{2020}\natexlab{b}.
\newblock \showarticletitle{Codimensional incremental potential contact}.
\newblock \bibinfo{journal}{\emph{arXiv preprint arXiv:2012.04457}}
  (\bibinfo{year}{2020}).
\newblock


\bibitem[Li et~al\mbox{.}(2023)]%
        {li2023subspace}
\bibfield{author}{\bibinfo{person}{Xuan Li}, \bibinfo{person}{Yu Fang},
  \bibinfo{person}{Lei Lan}, \bibinfo{person}{Huamin Wang},
  \bibinfo{person}{Yin Yang}, \bibinfo{person}{Minchen Li}, {and}
  \bibinfo{person}{Chenfanfu Jiang}.} \bibinfo{year}{2023}\natexlab{}.
\newblock \showarticletitle{Subspace-preconditioned gpu projective dynamics
  with contact for cloth simulation}. In \bibinfo{booktitle}{\emph{SIGGRAPH
  Asia 2023 Conference Papers}}. \bibinfo{pages}{1--12}.
\newblock


\bibitem[Liu et~al\mbox{.}(2016)]%
        {liu2016scalable}
\bibfield{author}{\bibinfo{person}{Haixiang Liu}, \bibinfo{person}{Nathan
  Mitchell}, \bibinfo{person}{Mridul Aanjaneya}, {and}
  \bibinfo{person}{Eftychios Sifakis}.} \bibinfo{year}{2016}\natexlab{}.
\newblock \showarticletitle{A scalable schur-complement fluids solver for
  heterogeneous compute platforms}.
\newblock \bibinfo{journal}{\emph{ACM Transactions on Graphics (TOG)}}
  \bibinfo{volume}{35}, \bibinfo{number}{6} (\bibinfo{year}{2016}),
  \bibinfo{pages}{1--12}.
\newblock


\bibitem[Long et~al\mbox{.}(2025)]%
        {long2025survey}
\bibfield{author}{\bibinfo{person}{Xiaoxiao Long}, \bibinfo{person}{Qingrui
  Zhao}, \bibinfo{person}{Kaiwen Zhang}, \bibinfo{person}{Zihao Zhang},
  \bibinfo{person}{Dingrui Wang}, \bibinfo{person}{Yumeng Liu},
  \bibinfo{person}{Zhengjie Shu}, \bibinfo{person}{Yi Lu},
  \bibinfo{person}{Shouzheng Wang}, \bibinfo{person}{Xinzhe Wei},
  \bibinfo{person}{Wei Li}, \bibinfo{person}{Wei Yin}, \bibinfo{person}{Yao
  Yao}, \bibinfo{person}{Jia Pan}, \bibinfo{person}{Qiu Shen},
  \bibinfo{person}{Ruigang Yang}, \bibinfo{person}{Xun Cao}, {and}
  \bibinfo{person}{Qionghai Dai}.} \bibinfo{year}{2025}\natexlab{}.
\newblock \showarticletitle{A Survey: Learning Embodied Intelligence from
  Physical Simulators and World Models}.
\newblock  (\bibinfo{date}{07} \bibinfo{year}{2025}).
\newblock
\href{https://doi.org/10.48550/arXiv.2507.00917}{doi:\nolinkurl{10.48550/arXiv.2507.00917}}


\bibitem[Low et~al\mbox{.}(2012)]%
        {low2012distributed}
\bibfield{author}{\bibinfo{person}{Yucheng Low}, \bibinfo{person}{Joseph
  Gonzalez}, \bibinfo{person}{Aapo Kyrola}, \bibinfo{person}{Danny Bickson},
  \bibinfo{person}{Carlos Guestrin}, {and} \bibinfo{person}{Joseph~M
  Hellerstein}.} \bibinfo{year}{2012}\natexlab{}.
\newblock \showarticletitle{Distributed graphlab: A framework for machine
  learning in the cloud}.
\newblock \bibinfo{journal}{\emph{arXiv preprint arXiv:1204.6078}}
  (\bibinfo{year}{2012}).
\newblock


\bibitem[Mashayekhi et~al\mbox{.}(2018)]%
        {mashayekhi2018automatically}
\bibfield{author}{\bibinfo{person}{Omid Mashayekhi}, \bibinfo{person}{Chinmayee
  Shah}, \bibinfo{person}{Hang Qu}, \bibinfo{person}{Andrew Lim}, {and}
  \bibinfo{person}{Philip Levis}.} \bibinfo{year}{2018}\natexlab{}.
\newblock \showarticletitle{Automatically distributing eulerian and hybrid
  fluid simulations in the cloud}.
\newblock \bibinfo{journal}{\emph{ACM Transactions on Graphics (TOG)}}
  \bibinfo{volume}{37}, \bibinfo{number}{2} (\bibinfo{year}{2018}),
  \bibinfo{pages}{1--14}.
\newblock


\bibitem[Nedic and Ozdaglar(2009)]%
        {nedic2009distributed}
\bibfield{author}{\bibinfo{person}{Angelia Nedic} {and} \bibinfo{person}{Asuman
  Ozdaglar}.} \bibinfo{year}{2009}\natexlab{}.
\newblock \showarticletitle{Distributed subgradient methods for multi-agent
  optimization}.
\newblock \bibinfo{journal}{\emph{IEEE Transactions on automatic control}}
  \bibinfo{volume}{54}, \bibinfo{number}{1} (\bibinfo{year}{2009}),
  \bibinfo{pages}{48--61}.
\newblock


\bibitem[Ouyang et~al\mbox{.}(2020)]%
        {ouyang2020anderson}
\bibfield{author}{\bibinfo{person}{Wenqing Ouyang}, \bibinfo{person}{Yue Peng},
  \bibinfo{person}{Yuxin Yao}, \bibinfo{person}{Juyong Zhang}, {and}
  \bibinfo{person}{Bailin Deng}.} \bibinfo{year}{2020}\natexlab{}.
\newblock \showarticletitle{Anderson acceleration for nonconvex ADMM based on
  Douglas-Rachford splitting}. In \bibinfo{booktitle}{\emph{Computer Graphics
  Forum}}, Vol.~\bibinfo{volume}{39}. Wiley Online Library,
  \bibinfo{pages}{221--239}.
\newblock


\bibitem[Overby et~al\mbox{.}(2017)]%
        {overby2017admm}
\bibfield{author}{\bibinfo{person}{Matthew Overby}, \bibinfo{person}{George~E.
  Brown}, \bibinfo{person}{Jie Li}, {and} \bibinfo{person}{Rahul Narain}.}
  \bibinfo{year}{2017}\natexlab{}.
\newblock \showarticletitle{ADMM $\supseteq$ Projective Dynamics: Fast
  Simulation of Hyperelastic Models with Dynamic Constraints}.
\newblock \bibinfo{journal}{\emph{IEEE Transactions on Visualization and
  Computer Graphics}} \bibinfo{volume}{23}, \bibinfo{number}{10}
  (\bibinfo{year}{2017}), \bibinfo{pages}{2222--2234}.
\newblock
\href{https://doi.org/10.1109/TVCG.2017.2730875}{doi:\nolinkurl{10.1109/TVCG.2017.2730875}}


\bibitem[Palomar and Chiang(2006)]%
        {palomar2006tutorial}
\bibfield{author}{\bibinfo{person}{Daniel~P{\'e}rez Palomar} {and}
  \bibinfo{person}{Mung Chiang}.} \bibinfo{year}{2006}\natexlab{}.
\newblock \showarticletitle{A tutorial on decomposition methods for network
  utility maximization}.
\newblock \bibinfo{journal}{\emph{IEEE Journal on Selected Areas in
  Communications}} \bibinfo{volume}{24}, \bibinfo{number}{8}
  (\bibinfo{year}{2006}), \bibinfo{pages}{1439--1451}.
\newblock


\bibitem[Pan and Manocha(2017)]%
        {pan2017efficient}
\bibfield{author}{\bibinfo{person}{Zherong Pan} {and} \bibinfo{person}{Dinesh
  Manocha}.} \bibinfo{year}{2017}\natexlab{}.
\newblock \showarticletitle{Efficient solver for spacetime control of smoke}.
\newblock \bibinfo{journal}{\emph{ACM Transactions on Graphics (ToG)}}
  \bibinfo{volume}{36}, \bibinfo{number}{4} (\bibinfo{year}{2017}),
  \bibinfo{pages}{1}.
\newblock


\bibitem[Peng et~al\mbox{.}(2018)]%
        {peng2018Anderson}
\bibfield{author}{\bibinfo{person}{Yue Peng}, \bibinfo{person}{Bailin Deng},
  \bibinfo{person}{Juyong Zhang}, \bibinfo{person}{Fanyu Geng},
  \bibinfo{person}{Wenjie Qin}, {and} \bibinfo{person}{Ligang Liu}.}
  \bibinfo{year}{2018}\natexlab{}.
\newblock \showarticletitle{Anderson Acceleration for Geometry Optimization and
  Physics Simulation}.
\newblock \bibinfo{journal}{\emph{ACM Transactions on Graphics}}
  \bibinfo{volume}{37}, \bibinfo{number}{4CD} (\bibinfo{year}{2018}),
  \bibinfo{pages}{42.1--42.14}.
\newblock


\bibitem[Qiu et~al\mbox{.}(2023)]%
        {qiu2023sparse}
\bibfield{author}{\bibinfo{person}{Yuxing Qiu}, \bibinfo{person}{Samuel~Temple
  Reeve}, \bibinfo{person}{Minchen Li}, \bibinfo{person}{Yin Yang},
  \bibinfo{person}{Stuart~Ryan Slattery}, {and} \bibinfo{person}{Chenfanfu
  Jiang}.} \bibinfo{year}{2023}\natexlab{}.
\newblock \showarticletitle{A sparse distributed gigascale resolution material
  point method}.
\newblock \bibinfo{journal}{\emph{ACM Transactions on Graphics}}
  \bibinfo{volume}{42}, \bibinfo{number}{2} (\bibinfo{year}{2023}),
  \bibinfo{pages}{1--21}.
\newblock


\bibitem[Rustico et~al\mbox{.}(2012)]%
        {rustico2012advances}
\bibfield{author}{\bibinfo{person}{Eugenio Rustico}, \bibinfo{person}{Giuseppe
  Bilotta}, \bibinfo{person}{Alexis Herault}, \bibinfo{person}{Ciro Del~Negro},
  {and} \bibinfo{person}{Giovanni Gallo}.} \bibinfo{year}{2012}\natexlab{}.
\newblock \showarticletitle{Advances in multi-GPU smoothed particle
  hydrodynamics simulations}.
\newblock \bibinfo{journal}{\emph{IEEE Transactions on Parallel and Distributed
  Systems}} \bibinfo{volume}{25}, \bibinfo{number}{1} (\bibinfo{year}{2012}),
  \bibinfo{pages}{43--52}.
\newblock


\bibitem[Shi et~al\mbox{.}(2015)]%
        {shi2015extra}
\bibfield{author}{\bibinfo{person}{Wei Shi}, \bibinfo{person}{Qing Ling},
  \bibinfo{person}{Gang Wu}, {and} \bibinfo{person}{Wotao Yin}.}
  \bibinfo{year}{2015}\natexlab{}.
\newblock \showarticletitle{Extra: An exact first-order algorithm for
  decentralized consensus optimization}.
\newblock \bibinfo{journal}{\emph{SIAM Journal on Optimization}}
  \bibinfo{volume}{25}, \bibinfo{number}{2} (\bibinfo{year}{2015}),
  \bibinfo{pages}{944--966}.
\newblock


\bibitem[Strm and Hgglund(1995)]%
        {1995PID}
\bibfield{author}{\bibinfo{person}{Karl~Johan Strm} {and} \bibinfo{person}{Tore
  Hgglund}.} \bibinfo{year}{1995}\natexlab{}.
\newblock \showarticletitle{PID controllers: Theory, Design and Tuning}.
\newblock \bibinfo{journal}{\emph{instrument society of america research
  triangle park nc}} (\bibinfo{year}{1995}).
\newblock


\bibitem[Tasora et~al\mbox{.}(2021)]%
        {tasora2021solving}
\bibfield{author}{\bibinfo{person}{Alessandro Tasora}, \bibinfo{person}{Dario
  Mangoni}, \bibinfo{person}{Simone Benatti}, {and} \bibinfo{person}{Rinaldo
  Garziera}.} \bibinfo{year}{2021}\natexlab{}.
\newblock \showarticletitle{Solving variational inequalities and cone
  complementarity problems in nonsmooth dynamics using the alternating
  direction method of multipliers}.
\newblock \bibinfo{journal}{\emph{Internat. J. Numer. Methods Engrg.}}
  \bibinfo{volume}{122}, \bibinfo{number}{16} (\bibinfo{year}{2021}),
  \bibinfo{pages}{4093--4113}.
\newblock


\bibitem[Tonge et~al\mbox{.}(2012)]%
        {tonge2012mass}
\bibfield{author}{\bibinfo{person}{Richard Tonge}, \bibinfo{person}{Feodor
  Benevolenski}, {and} \bibinfo{person}{Andrey Voroshilov}.}
  \bibinfo{year}{2012}\natexlab{}.
\newblock \showarticletitle{Mass splitting for jitter-free parallel rigid body
  simulation}.
\newblock \bibinfo{journal}{\emph{ACM Transactions on Graphics (TOG)}}
  \bibinfo{volume}{31}, \bibinfo{number}{4} (\bibinfo{year}{2012}),
  \bibinfo{pages}{1--8}.
\newblock


\bibitem[Wang et~al\mbox{.}(2023)]%
        {wang2023fast}
\bibfield{author}{\bibinfo{person}{Tianyu Wang}, \bibinfo{person}{Jiong Chen},
  \bibinfo{person}{Dongping Li}, \bibinfo{person}{Xiaowei Liu},
  \bibinfo{person}{Huamin Wang}, {and} \bibinfo{person}{Kun Zhou}.}
  \bibinfo{year}{2023}\natexlab{}.
\newblock \showarticletitle{Fast GPU-based two-way continuous collision
  handling}.
\newblock \bibinfo{journal}{\emph{ACM Transactions on Graphics}}
  \bibinfo{volume}{42}, \bibinfo{number}{5} (\bibinfo{year}{2023}),
  \bibinfo{pages}{1--15}.
\newblock


\bibitem[Wang et~al\mbox{.}(2020)]%
        {wang2020massively}
\bibfield{author}{\bibinfo{person}{Xinlei Wang}, \bibinfo{person}{Yuxing Qiu},
  \bibinfo{person}{Stuart~R Slattery}, \bibinfo{person}{Yu Fang},
  \bibinfo{person}{Minchen Li}, \bibinfo{person}{Song-Chun Zhu},
  \bibinfo{person}{Yixin Zhu}, \bibinfo{person}{Min Tang},
  \bibinfo{person}{Dinesh Manocha}, {and} \bibinfo{person}{Chenfanfu Jiang}.}
  \bibinfo{year}{2020}\natexlab{}.
\newblock \showarticletitle{A massively parallel and scalable multi-GPU
  material point method}.
\newblock \bibinfo{journal}{\emph{ACM Transactions on Graphics (TOG)}}
  \bibinfo{volume}{39}, \bibinfo{number}{4} (\bibinfo{year}{2020}),
  \bibinfo{pages}{30--1}.
\newblock


\bibitem[Xie et~al\mbox{.}(2023)]%
        {xie2023contact}
\bibfield{author}{\bibinfo{person}{Tianyi Xie}, \bibinfo{person}{Minchen Li},
  \bibinfo{person}{Yin Yang}, {and} \bibinfo{person}{Chenfanfu Jiang}.}
  \bibinfo{year}{2023}\natexlab{}.
\newblock \showarticletitle{A contact proxy splitting method for Lagrangian
  solid-fluid coupling}.
\newblock \bibinfo{journal}{\emph{ACM Transactions on Graphics (TOG)}}
  \bibinfo{volume}{42}, \bibinfo{number}{4} (\bibinfo{year}{2023}),
  \bibinfo{pages}{1--14}.
\newblock


\bibitem[Yang et~al\mbox{.}(2022)]%
        {yang2022survey}
\bibfield{author}{\bibinfo{person}{Yu Yang}, \bibinfo{person}{Xiaohong Guan},
  \bibinfo{person}{Qing-Shan Jia}, \bibinfo{person}{Liang Yu},
  \bibinfo{person}{Bolun Xu}, {and} \bibinfo{person}{Costas~J Spanos}.}
  \bibinfo{year}{2022}\natexlab{}.
\newblock \showarticletitle{A survey of ADMM variants for distributed
  optimization: Problems, algorithms and features}.
\newblock \bibinfo{journal}{\emph{arXiv preprint arXiv:2208.03700}}
  (\bibinfo{year}{2022}).
\newblock


\bibitem[Yuan et~al\mbox{.}(2018)]%
        {yuan2018exact}
\bibfield{author}{\bibinfo{person}{Kun Yuan}, \bibinfo{person}{Bicheng Ying},
  \bibinfo{person}{Xiaochuan Zhao}, {and} \bibinfo{person}{Ali~H Sayed}.}
  \bibinfo{year}{2018}\natexlab{}.
\newblock \showarticletitle{Exact diffusion for distributed optimization and
  learning—Part I: Algorithm development}.
\newblock \bibinfo{journal}{\emph{IEEE Transactions on Signal Processing}}
  \bibinfo{volume}{67}, \bibinfo{number}{3} (\bibinfo{year}{2018}),
  \bibinfo{pages}{708--723}.
\newblock


\bibitem[Zaharia et~al\mbox{.}(2012)]%
        {zaharia2012resilient}
\bibfield{author}{\bibinfo{person}{Matei Zaharia}, \bibinfo{person}{Mosharaf
  Chowdhury}, \bibinfo{person}{Tathagata Das}, \bibinfo{person}{Ankur Dave},
  \bibinfo{person}{Justin Ma}, \bibinfo{person}{Murphy McCauly},
  \bibinfo{person}{Michael~J Franklin}, \bibinfo{person}{Scott Shenker}, {and}
  \bibinfo{person}{Ion Stoica}.} \bibinfo{year}{2012}\natexlab{}.
\newblock \showarticletitle{Resilient distributed datasets: A
  $\{$Fault-Tolerant$\}$ abstraction for $\{$In-Memory$\}$ cluster computing}.
  In \bibinfo{booktitle}{\emph{9th USENIX symposium on networked systems design
  and implementation (NSDI 12)}}. \bibinfo{pages}{15--28}.
\newblock


\bibitem[Zhang et~al\mbox{.}(2019)]%
        {Zhang2019Accelerating}
\bibfield{author}{\bibinfo{person}{Juyong Zhang}, \bibinfo{person}{Yue Peng},
  \bibinfo{person}{Wenqing Ouyang}, {and} \bibinfo{person}{Bailin Deng}.}
  \bibinfo{year}{2019}\natexlab{}.
\newblock \showarticletitle{Accelerating ADMM for efficient simulation and
  optimization}.
\newblock \bibinfo{journal}{\emph{ACM Transactions on Graphics (TOG)}}
  \bibinfo{volume}{38}, \bibinfo{number}{6} (\bibinfo{year}{2019}),
  \bibinfo{pages}{1--21}.
\newblock


\bibitem[Zheng et~al\mbox{.}(2025)]%
        {zheng2025robust}
\bibfield{author}{\bibinfo{person}{Juntian Zheng}, \bibinfo{person}{Zhaofeng
  Luo}, {and} \bibinfo{person}{Minchen Li}.} \bibinfo{year}{2025}\natexlab{}.
\newblock \showarticletitle{Robust and Efficient Penetration-Free
  Elastodynamics without Barriers}.
\newblock \bibinfo{journal}{\emph{arXiv preprint arXiv:2512.12151}}
  (\bibinfo{year}{2025}).
\newblock


\end{thebibliography}


\newpage


\clearpage

\clearpage

\setcounter{figure}{9}
\aptLtoX{\begin{figure}[t!]
  \centering
\includegraphics[width=\textwidth,height=0.95\textheight,keepaspectratio]{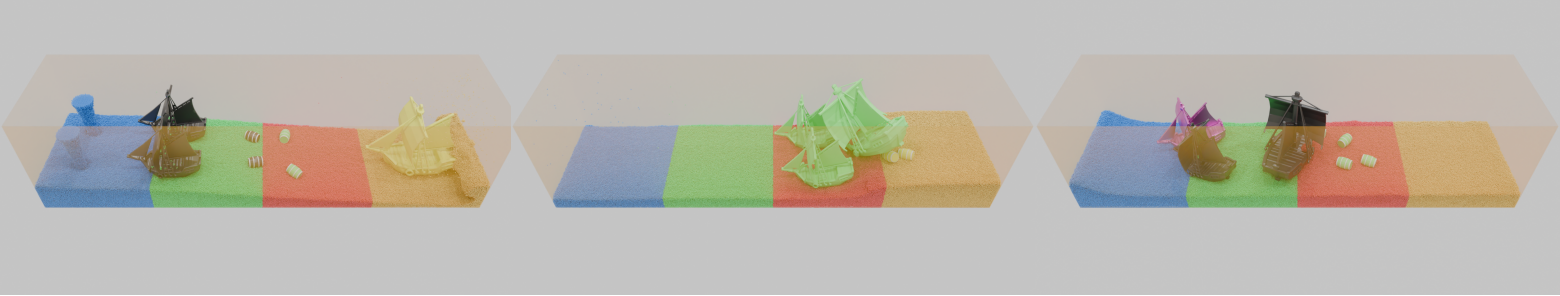}
  \caption{\emph{Solid-fluid coupling.} 
Alongside distributed ABD, our system also supports distributed fluid simulation and fluid–solid coupling. In this demo, four workers run distributed IISPH with 1M fluid particles and an ABD solid scene of 7 bodies totaling 220K triangles, with impulse-based two-way solid–fluid coupling. These results show that our multi-node architecture can support coupled multiphysics within a unified distributed framework, paving the way towards large-scale multiphysics simulation on multi-node systems.
  }
  \label{fig:demo_coupling}
  \end{figure}
  
  \begin{figure}
    \centering
    \includegraphics[width=0.85\linewidth]{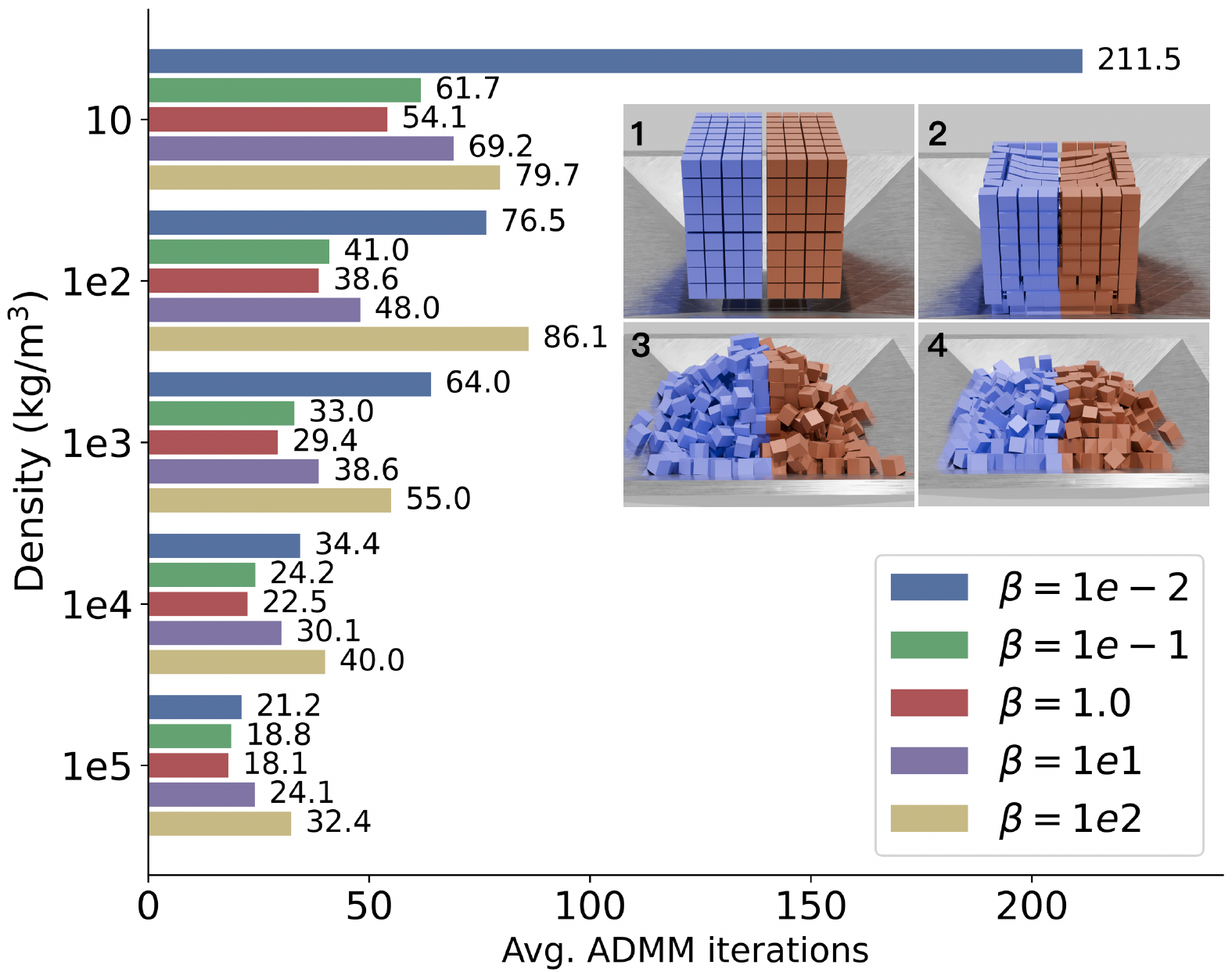}
    \vspace{-6pt}
    \caption{Mass scaling for penalty initialization. We vary density to scale body masses and report average ADMM iterations. Initializing $\rho$ with body mass reduces iteration counts across all settings, implying faster convergence.}
    \label{fig:mass-exp}
  \end{figure}
  
  \begin{figure}
    \centering
    \includegraphics[width=0.95\linewidth]{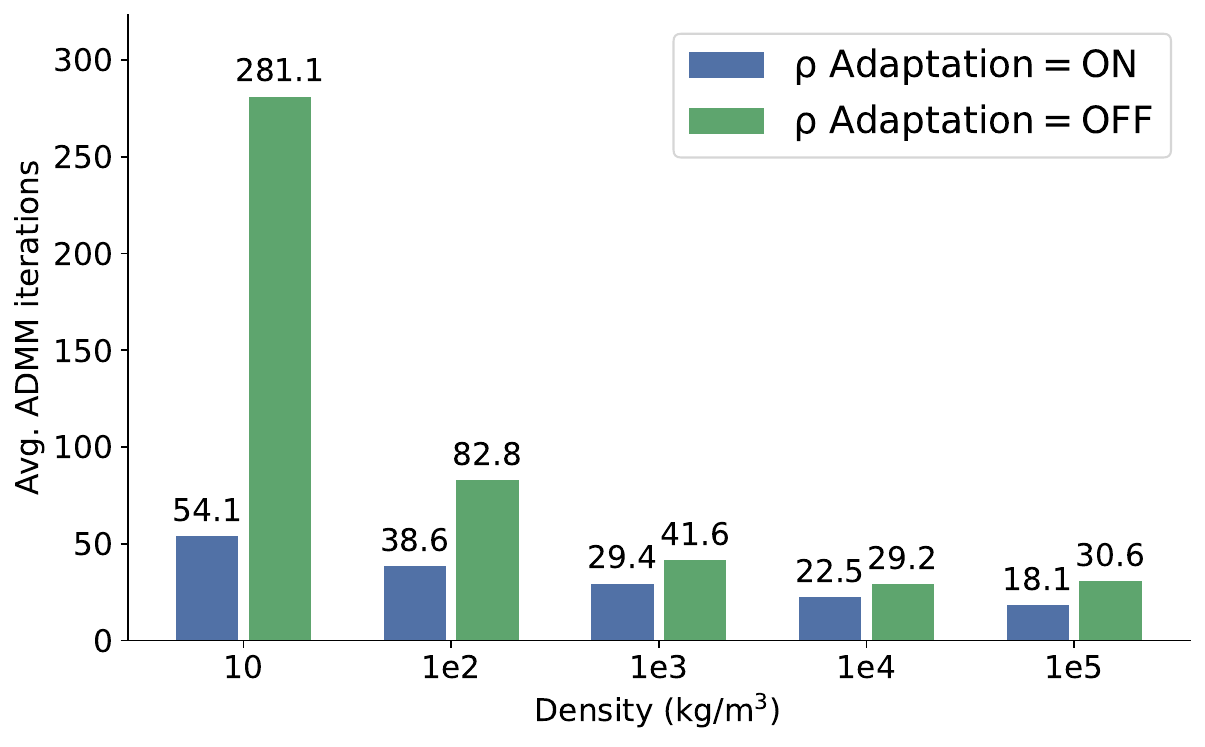}
    \caption{Ablation of residual-driven $\rho$ adaptation. For five mass settings, we report the average ADMM iterations with and without dynamic $\rho$ updates.}
    \label{fig:rho_adap}
   \end{figure}

  \begin{figure} 
  \centering
      \includegraphics[width=0.95\linewidth]{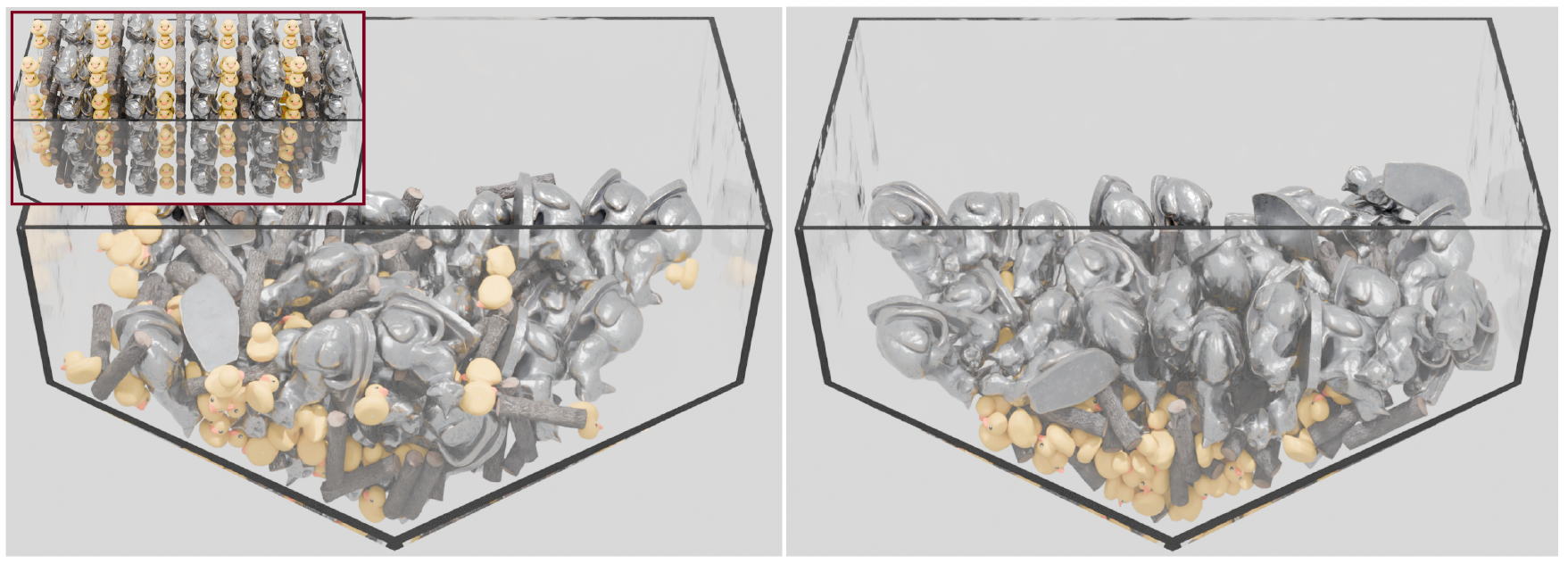}
\caption{\changed{}{Experimental scene for evaluating the effect of mass and stiffness heterogeneity. The inset at the upper left shows the initial configuration. Ducks, wooden logs, and metallic cats correspond to $(m,E)=(1,10^8)$, $(100,10^{10})$, and $(10^4,10^{12})$, respectively. The left panel shows the final state of the heterogeneous setting, while the right panel shows the corresponding uniform setting with $(m,E)=(100,10^{10})$ for all objects. Both panels share the same geometry and initial layout.}}
      \label{fig:exp-vary-mass}
\end{figure}}{\begin{figure*}[t!]
  \centering
  \begin{minipage}[t]{0.95\textwidth}
    \centering
\includegraphics[width=\textwidth,height=0.95\textheight,keepaspectratio]{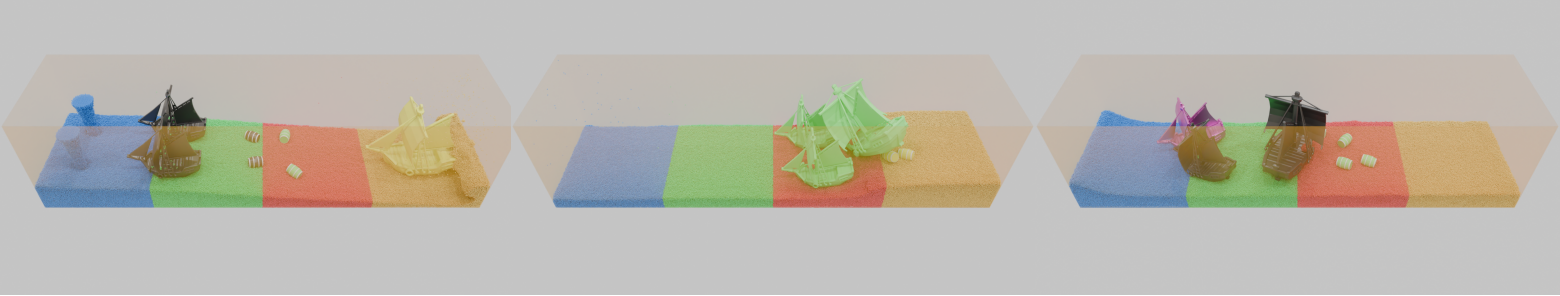}
  \caption{\emph{Solid-fluid coupling.} 
Alongside distributed ABD, our system also supports distributed fluid simulation and fluid–solid coupling. In this demo, four workers run distributed IISPH with 1M fluid particles and an ABD solid scene of 7 bodies totaling 220K triangles, with impulse-based two-way solid–fluid coupling. These results show that our multi-node architecture can support coupled multiphysics within a unified distributed framework, paving the way towards large-scale multiphysics simulation on multi-node systems.
  }
  \label{fig:demo_coupling}
  \end{minipage}
  
  \begin{minipage}[t]{0.45\textwidth}
    \centering
    \includegraphics[width=0.85\linewidth]{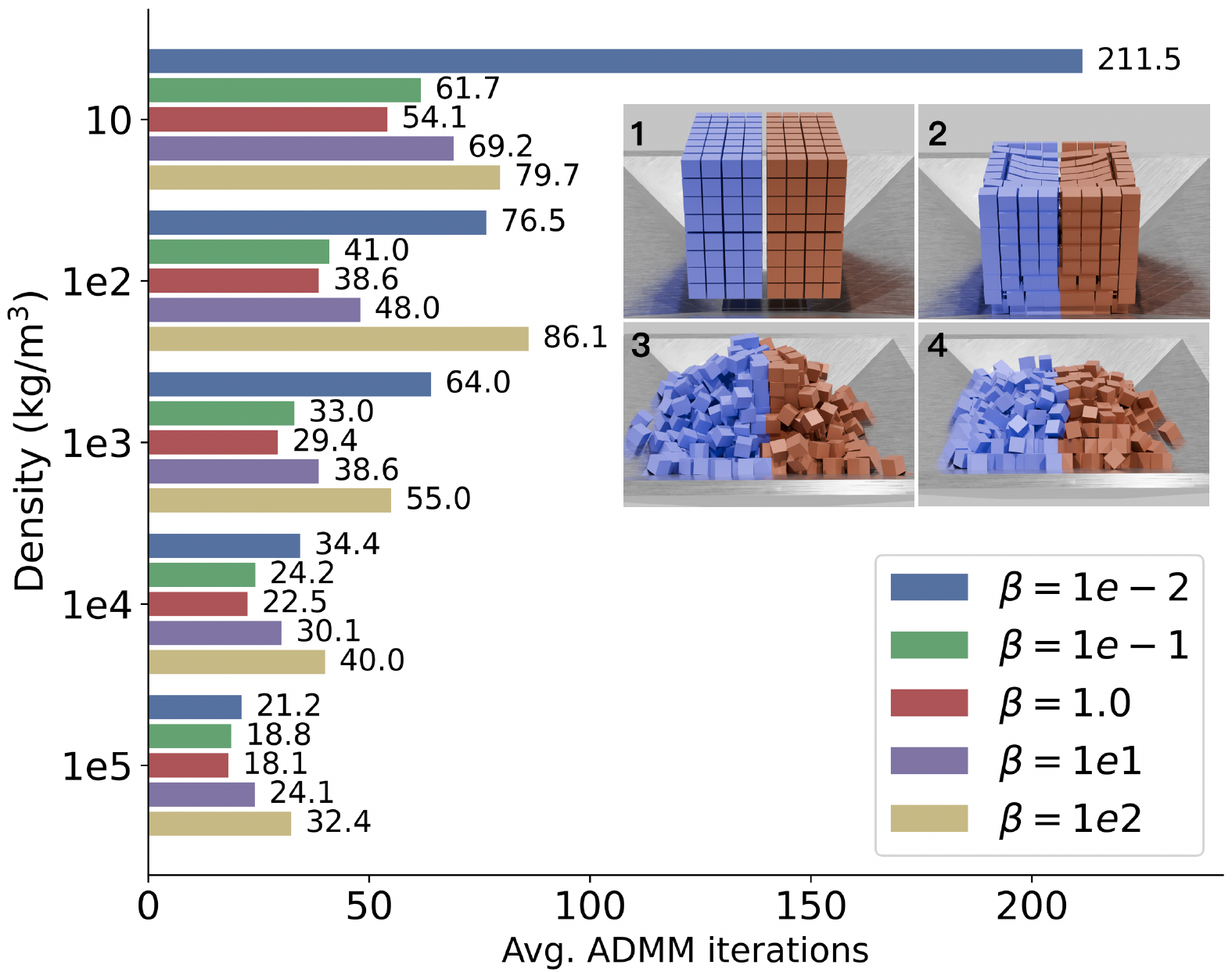}
    \vspace{-6pt}
    \caption{Mass scaling for penalty initialization. We vary density to scale body masses and report average ADMM iterations. Initializing $\rho$ with body mass reduces iteration counts across all settings, implying faster convergence.}
    \label{fig:mass-exp}
  \end{minipage}
  \hfill 
  \begin{minipage}[t]{0.45\textwidth}
    \centering
    \includegraphics[width=0.95\linewidth]{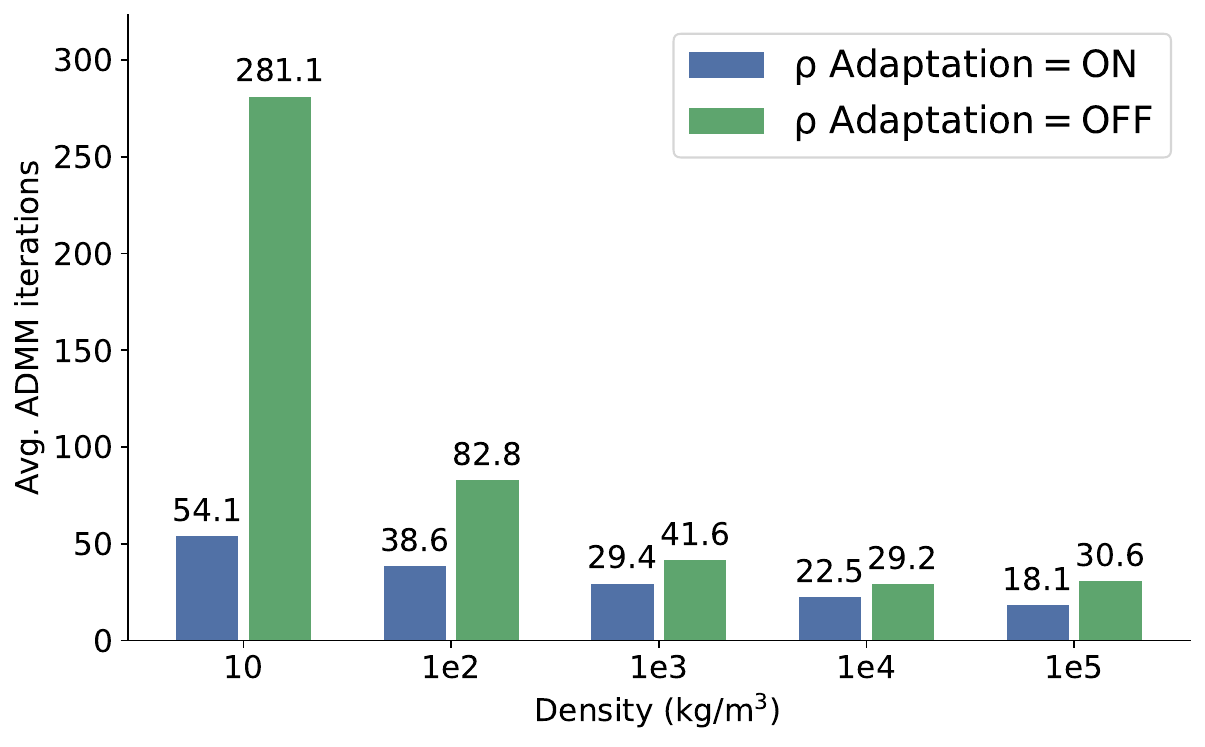}
    \caption{Ablation of residual-driven $\rho$ adaptation. For five mass settings, we report the average ADMM iterations with and without dynamic $\rho$ updates.}
    \label{fig:rho_adap}
   \end{minipage}

  \begin{minipage}[t]{0.45\textwidth} 
  \centering
      \includegraphics[width=0.95\linewidth]{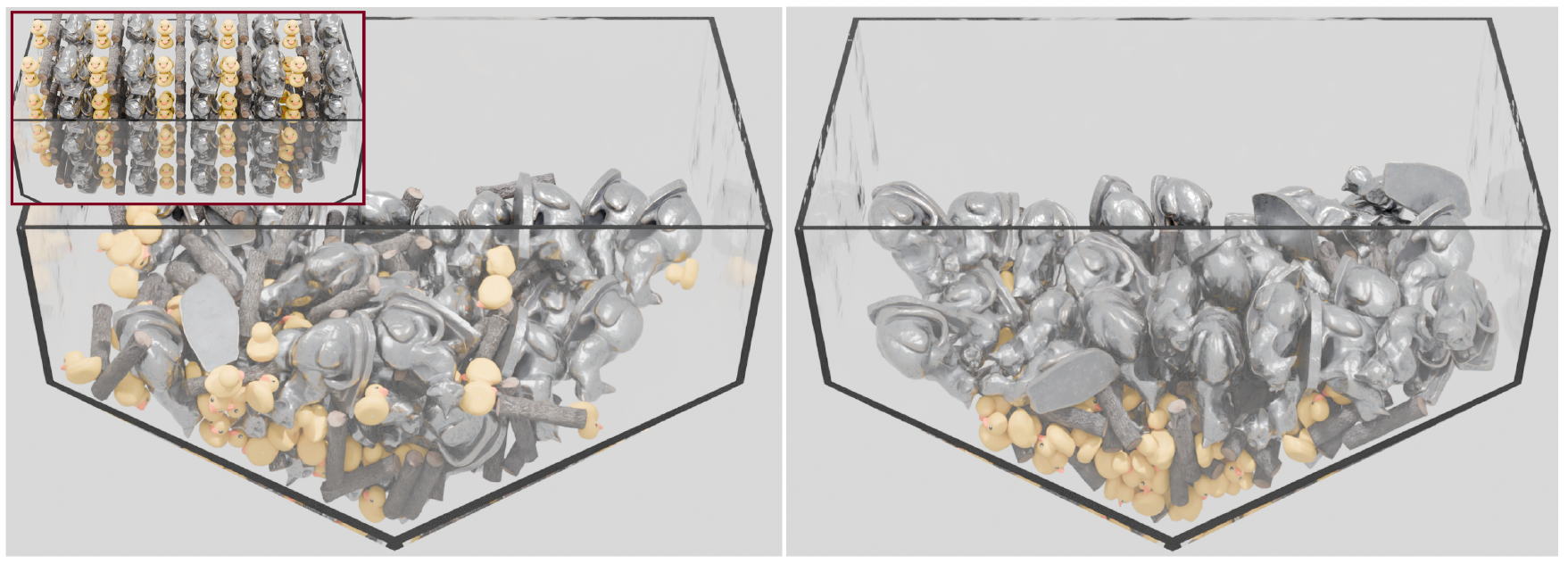}
\caption{\changed{}{Experimental scene for evaluating the effect of mass and stiffness heterogeneity. The inset at the upper left shows the initial configuration. Ducks, wooden logs, and metallic cats correspond to $(m,E)=(1,10^8)$, $(100,10^{10})$, and $(10^4,10^{12})$, respectively. The left panel shows the final state of the heterogeneous setting, while the right panel shows the corresponding uniform setting with $(m,E)=(100,10^{10})$ for all objects. Both panels share the same geometry and initial layout.}}
      \label{fig:exp-vary-mass}
  \end{minipage}
  \hfill
  \begin{minipage}[t]{0.45\textwidth} 
  \end{minipage}
\end{figure*}}




\end{document}